\pdfoutput=1
\documentclass[cernpreprint,atlasdraft=false,texlive=2016,backref= false,UKenglish,english,texmf]{atlasdoc}
 
\usepackage{atlaspackage}
\usepackage{atlasbiblatex}
 
\usepackage{atlasphysics}
 
\graphicspath{{logos/}{figures/}}
 
\usepackage{ANA-SUSY-2018-15-PAPER-defs}
\usepackage{multirow}

\AtlasTitle{A search for the decays of stopped long-lived particles at $\sqrt{s}=13$~\TeV\ with the ATLAS detector}
 
\AtlasAbstract{
A search for long-lived particles, which have come to rest within the ATLAS detector, is presented.
The subsequent decays of these particles can produce high-momentum jets, resulting in large out-of-time energy deposits in the ATLAS calorimeters.
These decays are detected using data collected during periods in the LHC bunch structure when collisions are absent.
The analysed dataset is composed of events from proton--proton collisions produced by the Large Hadron Collider at a centre-of-mass energy of $\sqrt{s}=13$~\TeV\ and recorded by
the ATLAS experiment during 2017 and 2018.
The dataset used for this search corresponds to a total live time of 579~hours.
The results of this search are used to derive lower limits on the mass of gluino $R$-hadrons, assuming a branching fraction
$\mathcal{B}(\tilde{g}\rightarrow \qqbar \tilde{\chi}_1^0)=100$\%, with masses of up to $1.4$~\TeV\ excluded for gluino lifetimes of $10^{-5}$ to $10^3$~s.
}
 
\author{The ATLAS Collaboration}
 
\PreprintIdNumber{CERN-EP-2021-041}

\AtlasJournal{JHEP}
\AtlasJournalRef{JHEP 07 (2021) 173}
\AtlasDOI{10.1007/JHEP07(2021)173}
\hypersetup{pdftitle={ATLAS document},pdfauthor={The ATLAS Collaboration}}
 
\begin{document}
 
\maketitle

\section{Introduction}
\label{sec:intro}
 
This paper presents a search for long-lived particles which have come to rest within the ATLAS calorimeters and decay at a later time when no proton--proton ($pp$) collisions occur.
 
Although absent in the Standard Model (SM), exotic metastable particles are featured in many beyond-the-SM (BSM) theories.
These include R-parity-conserving supersymmetry (SUSY)~\cite{Farrar:1978xj,Golfand:1971iw,Volkov:1973ix,Wess:1974tw,Wess:1974jb,Ferrara:1974pu,Salam:1974ig} models such as split-SUSY~\cite{Giudice:2004tc,ArkaniHamed:2004fb} and gauge-mediated SUSY breaking~\cite{Dine:1981gu,AlvarezGaume:1981wy,Nappi:1982hm}, as well as other scenarios such as universal extra dimensions~\cite{Appelquist:2000nn,Cheng:2002ab}.
In the example of split-SUSY, it is assumed that the naturalness problem is solved via a small amount of fine-tuning, and gauge coupling unification and dark matter are used to guide the theory.
Here, SUSY-breaking can occur at a very high energy scale, leading to correspondingly heavy squarks and sleptons.
This results in a suppression of gluino decay via a heavy off-shell squark, creating the possibility for the gluino to acquire a non-negligible mean proper lifetime \taugl.
 
If long-lived gluinos are produced by proton collisions at the Large Hadron Collider (LHC), these strongly produced SUSY particles would hadronise with SM quarks and gluons, forming new composite states known as \textit{\rhadrons}~\cite{Farrar:1978xj}. This scenario was chosen as the benchmark for this search.
The constituent gluino or squark can be regarded as a heavy particle surrounded by a cloud of interacting, coloured SM particles. 
As the \rhadron\ traverses the detector it interacts with the detector material via the exchange of constituent partons, thereby altering its composition.
During this process the \rhadron\ can flip between neutral, charged or even doubly charged states.
This search is robust against these charge flips.
Since in most models the lightest $R$-baryon state is charge neutral~\cite{ATL-PHYS-PUB-2019-019}, a significant fraction of \rhadrons\ that exit the ATLAS calorimeter would be charge neutral.
Detector-stable \rhadrons\ are therefore likely to not leave charged tracks in the muon spectrometer system.
 
Assuming that the gluinos are produced near threshold at the LHC, the \rhadrons\ they form are expected to be slow-moving,
and some fraction will lose sufficient momentum while traversing the detector to come to rest.
In such cases a stopped gluino could decay significantly later than the bunch crossing in which it was produced, depending on \taugl, leaving a significant energy deposit within the ATLAS calorimeter system.
 
This search makes use of data collected in so-called \textit{empty} bunch crossings, where the proton buckets in the crossing beams are unfilled, providing a clean environment in which to identify these delayed decays. This approach provides sensitivity to gluino lifetimes across several orders of magnitude, in the range of $10^{-5}$ to $10^{3}$~s. In the absence of proton collisions, events containing muons from cosmic rays or the interaction of beam protons with upstream collimators, residual gas within the beam pipe, or the beam pipe itself, become the dominant background.
 
Similar searches have been performed previously by the ATLAS and CMS collaborations at $\sqrt{s}=7$,
$8$ and $13$~\TeV~\cite{SUSY-2011-01,SUSY-2013-03,CMS-EXO-10-003,CMS-EXO-11-020,CMS-EXO-12-036,CMS-EXO-16-004}.
This analysis represents the first search for stopped long-lived particles at ATLAS using $\sqrt{s}=13$~\TeV\ $pp$ collision data,
with an integrated luminosity of 111~\ifb.
The analysis significantly expands the limits on such signatures given by previous ATLAS analyses, which exclude gluinos with mass $m(\tilde{g})<832$~\GeV\ for \taugl\ from $10^{-5}$ to $10^3$~s.
Novel strategies for the estimation of non-collision background processes have been developed,
making use of additional non-collision datasets collected by the ATLAS experiment to derive background templates.
 
 
\section{ATLAS detector}
\label{sec:detector}
 
\newcommand{\AtlasCoordFootnote}{
ATLAS uses a right-handed coordinate system with its origin at the nominal interaction point (IP)
in the centre of the detector and the \(z\)-axis along the beam pipe.
The \(x\)-axis points from the IP to the centre of the LHC ring,
and the \(y\)-axis points upwards.
Cylindrical coordinates \((r,\phi)\) are used in the transverse plane,
\(\phi\) being the azimuthal angle around the \(z\)-axis.
The pseudorapidity is defined in terms of the polar angle \(\theta\) as \(\eta = -\ln \tan(\theta/2)\).
Angular distance is measured in units of \(\Delta R \equiv \sqrt{(\Delta\eta)^{2} + (\Delta\phi)^{2}}\).}
 
The ATLAS detector~\cite{PERF-2007-01} at the LHC covers nearly the entire solid angle around the collision point.\footnote{\AtlasCoordFootnote}
It consists of an inner tracking detector surrounded by a thin superconducting solenoid, electromagnetic and hadronic calorimeters,
and a muon spectrometer incorporating three large superconducting toroidal magnets.
 
The inner-detector system is immersed in a \SI{2}{\tesla} axial magnetic field
and provides charged-particle tracking in the range \(|\eta| < 2.5\).
The high-granularity silicon pixel detector covers the vertex region and typically provides four measurements per track,
the first hit normally being in the insertable B-layer installed before Run~2~\cite{ATLAS-TDR-2010-19,PIX-2018-001}.
It is followed by the silicon microstrip tracker, which usually provides eight measurements per track.
These silicon detectors are complemented by the transition radiation tracker (TRT),
which enables radially extended track reconstruction up to \(|\eta| = 2.0\).
The TRT also provides electron identification information
based on the fraction of hits (typically 30 in total) above a higher energy-deposit threshold corresponding to transition radiation.
 
The calorimeter system covers the pseudorapidity range \(|\eta| < 4.9\).
Within the region \(|\eta|< 3.2\), electromagnetic calorimetry is provided by electromagnetic barrel and
endcap high-granularity lead/liquid-argon (LAr) calorimeters,
with an additional thin LAr presampler covering \(|\eta| < 1.8\)
to correct for energy loss in material upstream of the calorimeters.
Hadronic calorimetry is provided by the steel/scintillator-tile (Tile) calorimeter,
segmented into three barrel structures within \(|\eta| < 1.7\), and two copper/LAr hadronic endcap calorimeters (HEC).
The solid angle coverage is completed with forward copper/LAr and tungsten/LAr calorimeter modules
optimised for electromagnetic and hadronic measurements respectively.
 
The muon spectrometer (MS) comprises separate trigger and
high-precision tracking chambers measuring the deflection of muons in a magnetic field generated by the superconducting air-core toroids.
The field integral of the toroids ranges between \num{2.0} and \SI{6.0}{\tesla\metre}
across most of the detector.
A set of precision chambers covers the region \(|\eta| < 2.7\) with three layers of monitored drift tubes (MDTs),
complemented by cathode-strip chambers (CSCs) in the forward region, where the background is highest.
The muon trigger system covers the range \(|\eta| < 2.4\) with resistive-plate chambers in the barrel, and thin-gap chambers in the endcap regions.
Interesting events are selected to be recorded by the first-level (L1) trigger system implemented in custom hardware,
followed by selections made by algorithms implemented in software in the high-level trigger (HLT)~\cite{TRIG-2016-01}.
The L1 trigger accepts events from the \SI{40}{\MHz} bunch crossings at a rate below \SI{100}{\kHz},
which the HLT reduces in order to record events to disk at about \SI{1}{\kHz}.
 
 
\section{Dataset and reconstruction}
\label{sec:data}
 
During $pp$ collision data-taking the LHC circulates two counter-rotating proton beams constructed from bunches of protons ($\sim$$10^{11}$ protons per bunch).
The radio frequency (RF) cavities providing particle acceleration at the LHC operate at 400~MHz,
which corresponds to an RF bucket spacing of 2.5~ns.
A group of ten RF buckets are assigned a unique bunch-crossing identifier (BCID) within which only one RF bucket can contain a proton bunch, for each beam.
Proton bunches are spaced at 25~ns intervals, such that a filled RF bucket is separated from the next by at least 25~ns.
There are 3564 available BCIDs in which a filled bunch can reside around the LHC circumference, with each corresponding to a time window of 25~ns.
Following LHC injection not all BCIDs contain a filled bunch, with the number of unfilled bunches depending on the LHC filling scheme~\cite{Evans:2008zzb}.
 
To identify potential signals in the calorimeter that originate from delayed decays of stopped exotic particles,
this analysis makes use of events recorded during \textit{empty} bunch crossings (BXs), where the ten crossing RF buckets in each beam are unfilled.
Events selected for analysis are taken from empty BXs during stable beam periods, when data-taking for physics purposes is underway,
and are required to pass standard data-quality requirements~\cite{DAPR-2018-01}.
The use of empty bunches minimises detector activity from collision events, reducing contamination from SM processes in a bid to identify out-of-time decays.
As a result,
the main backgrounds for this search are from non-collision sources, which, to leading order, scale with \textit{live time} as opposed to the integrated luminosity of the proton--proton data-taking campaign.
Here `live time' refers to the total amount of time during which the trigger was able to select and accept signal-like events in empty BXs in the $\sqrt{s}=13$~\TeV\ $pp$ data-taking.
In contrast to the background, the number of signal events in a given dataset approximately scales with both the live time, governing the acceptance of potential signal events, and the integrated luminosity of the $pp$ collisions, which governs their production.

This analysis includes data collected during 2017, with an empty-BX live time of 298 hours, and during 2018, with an empty-BX live time of 281 hours.
These are the live times for this analysis after imposing requirements based on beam conditions, detector conditions and data quality.
The 2017 dataset corresponds to an integrated luminosity of $49.0$~\ifb\ of $\sqrt{s}=13$~\TeV\ $pp$ collisions delivered by the LHC to the ATLAS experiment,
while the 2018 dataset corresponds to $62.1$~\ifb.
The uncertainty in the 2017 (2018) integrated luminosity is 2.4\% (2.0\%)~\cite{ATLAS-CONF-2019-021}, obtained using the LUCID-2 detector~\cite{Avoni:2018iuv} for the primary luminosity measurements.
The analysis sensitivity is higher for the 2018 dataset,
which has the larger integrated luminosity and a slightly lower live time than the 2017 dataset.
The higher live time during 2017 results in a larger contribution from non-collision backgrounds,
while the lower integrated luminosity of the 2017 dataset leads to fewer expected signal events.
For 2015 and 2016 the even lower integrated luminosities and higher live times reduce the sensitivity of the analysis further,
motivating the use of the 2017 and 2018 datasets alone in this search.
The exception to this is a single dataset from 2016, taken during a period of time when there were no beams circulating within the LHC machine,
which is used to study backgrounds induced by cosmic rays (a background component referred to herein as `cosmics').

\subsection{Analysis data samples and triggers}
 
Dedicated triggers are used to select empty-BX events with significant calorimeter activity.
For the 2017 data, these triggers require that the empty bunch in each beam is separated from a filled bunch by at least five unfilled BCIDs both before and after.
This effectively enforces a $150$~ns buffer between the empty BXs considered in this analysis and a collision event.
Due to the introduction of additional trigger acceptance, in the 2018 dataset this buffer is reduced to $100$~ns before the considered BX,
but for BXs following the considered BX the buffer remains at $150$~ns.
`Signal-like' events forming the \textit{search sample} for this analysis are selected in empty BXs by a trigger requiring missing transverse momentum $\met >50$~\GeV\ and at least one jet with transverse momentum $\pt >55$~\GeV\ and $|\eta|<2.4$ at the HLT.
In order to be selected by this trigger the event must contain at least one jet with $\pt>30$~\GeV\ at L1.
To ensure the trigger is fully efficient relative to the analysis selection,
events selected for analysis are required to contain at least one jet with $\pt>90$~\GeV.

Additional datasets are used to study background processes.
Cosmic rays can result in energetic muons traversing the detector, which can in turn induce energetic jets in the calorimeters, presenting an important background for this search.
To obtain a pure sample of cosmic-ray events free of beam-induced backgrounds (BIB), a cosmic run (taken without beam in the LHC machine) recorded during 2016 provides a \textit{cosmic sample}.
Beam-induced backgrounds in this search arise as a result of stray protons interacting with material upstream from the detector,
producing energetic particles that traverse the detector in the horizontal plane.
Jets induced by energetic muons produced in these interactions present the other important background process for this search.
An additional \textit{BIB sample} for studying beam-induced backgrounds is taken from \textit{unpaired} BX data.
Here only one of the crossing LHC beams contains a filled bunch (with the corresponding bunch in the other beam remaining unfilled),
reducing contributions from beam--beam collisions.
Events are selected from these datasets using a trigger requiring the presence of a jet with \pt\ greater than either $12$~\GeV\ or $50$~\GeV\ at L1.
Because of its high rate, only a fraction of the events passing the $12$~\GeV\ trigger is recorded.
This trigger, therefore, is only used to select events in which the highest jet \pt\ is under $120$~\GeV.
The contribution from events selected by this trigger is scaled by the inverse of the corresponding fraction,
to ensure a smooth jet-\pt\ distribution when transitioning between events selected using the $12$~\GeV\ and $50$~\GeV\ L1 trigger
(the latter trigger is used to select events in which the highest jet \pt\ is ${\geq}120$~\GeV).
An additional \textit{cavern background sample} is formed using events recorded by a random trigger, which was active during empty BXs.
This is used to provide an unbiased sample from which to derive the amount of noise expected in the ATLAS detector due to effects such as activity from errant neutrons and photons commonly called `cavern background'.
A summary of the data samples used in this analysis is given in Table~\ref{tab:datasets}.
The same $\pt$ requirements are placed on each data sample used in the analysis in order to make the samples selected using different triggers consistent.
The exception to this is the cavern background sample, where an unbiased sample is obtained by making no kinematic selection.

\begin{table}
\begin{center}
\caption{Overview of the data samples used in this analysis. The kinematic requirements at the HLT or L1 trigger that events are required to pass in order to be recorded are indicated under `trigger requirements', with the corresponding requirements enforced at analysis level indicated under `offline requirements'.
No `bunch structure' is listed for the cosmic sample, since no beam was circulating in the LHC machine while these cosmic data were being collected.}
\vspace{0.2cm}
\begin{tabular}{llll}
\hline
\textbf{Data sample}                    & \textbf{Bunch}         & \textbf{Trigger}                  & \textbf{Offline}      \\
\textbf{(purpose)}                      & \textbf{structure}     & \textbf{requirements}             & \textbf{requirements} \\
\hline\hline
Search sample                  & Empty         & HLT jet \pt$>55$~\GeV    & Leading jet \pt$>90$~\GeV \\
&               & HLT \met$>50$~\GeV       & \\
&               & HLT jet $|\eta|<2.4$     & Leading jet $|\eta|<2.4$ \\
\hline
Cosmic sample                 & \multicolumn{1}{c}{$-$}           & L1 jet \pt$>12$~\GeV     & Leading jet \pt$>90$~\GeV \\
&               &                          & Leading jet $|\eta|<2.4$ \\
\hline
Beam-induced background sample & Unpaired      & L1 jet \pt$>12$~\GeV\ or & Leading jet \pt$>90$~\GeV \\
&               & L1 jet \pt$>50$~\GeV     & \\
&               &                          & Leading jet $|\eta|<2.4$ \\
\hline
Cavern background sample       & Empty         & Random                   & \multicolumn{1}{c}{$-$}\\
\hline
\end{tabular}
\label{tab:datasets}
\end{center}
\end{table}
 
\FloatBarrier
\subsection{Data reconstruction}
 
The signal decays targeted by this search would be expected to originate from within the detector material, and be displaced from the interaction point (IP).
The decay of a stationary long-lived particle would be isotropic, and so the decay activity would likely not point directly towards or away from the IP,
nor would resulting energy deposits necessarily be projective with respect to the expected positions of beam--beam collision vertices.
In order to increase the efficiency with which out-of-time and non-pointing signals can be reconstructed,
and to increase the identification (and subsequently rejection) efficiency of background events from cosmic rays,
a \textit{cosmic} reconstruction configuration~\cite{PERF-2010-02} is applied to reconstruct all data used in this search.
This configuration loosens the pointing and impact parameter requirements on muon segments~\cite{Nicolaidou:2010zz},
such that muon tracks are not required to point towards the IP.
Special timing procedures are applied to cosmic muon reconstruction in the MDT detectors to account for the different timing of the recorded hits relative to the nominal LHC bunch crossing.
These modifications increase the efficiency with which the upper muon segments produced by a cosmic-ray muon are reconstructed.
This is illustrated in Figure~\ref{fig:cosmic_display}, where both the upper and lower muon segments induced by a traversing cosmic-ray muon are reconstructed when using the cosmic reconstruction mode,
but when using a \textit{collision} reconstruction configuration only the lower segments are reconstructed.
 
\begin{figure}
\centering
\includegraphics[width=0.8\textwidth]{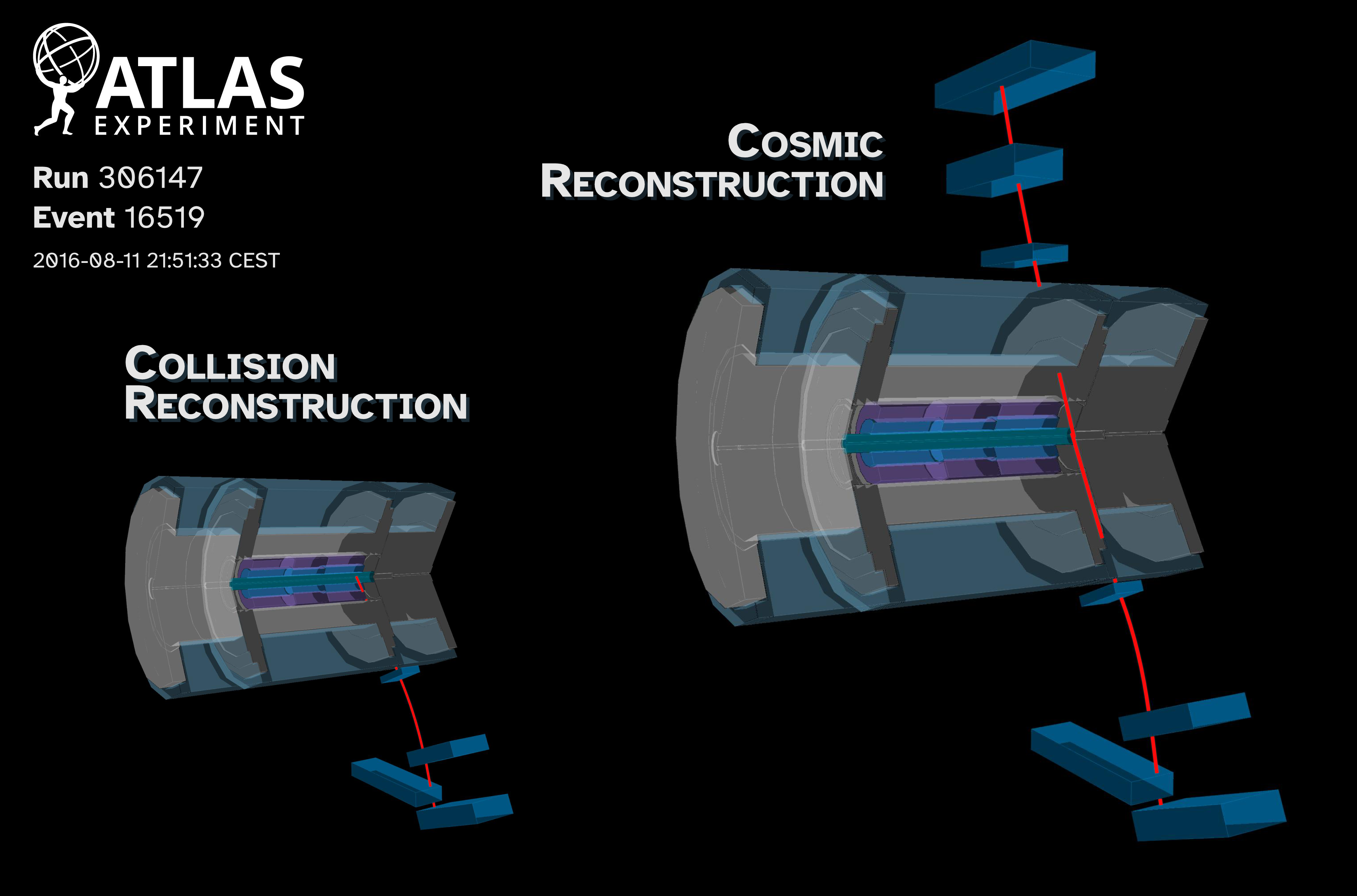}
\caption{Event display showing an event from the cosmic sample reconstructed using \textit{collision} reconstruction (left) and \textit{cosmic} reconstruction (right) configurations.
The red line represents the extrapolated muon path based on the hits induced in the MDTs (blue).
For readability, only the MDT chambers with associated hits are shown.
A cut-away view of the LAr and Tile calorimeters is included for orientation.
By using \textit{cosmic} reconstruction mode the muon-segment track reconstruction requirements are loosened, increasing the efficiency for identifying muon segments induced by cosmic rays.
With \textit{collision} reconstruction settings, only the lower half of the cosmic-ray muon track is reconstructed, whereas in \textit{cosmic} reconstruction mode the upper leg is reconstructed as well.}
\label{fig:cosmic_display}
\end{figure}

The LAr calorimeter reconstruction method remains unchanged when reconstructing the data in \textit{cosmic} mode,
while the Tile calorimeter reconstruction methods are altered to use an iterative method that more accurately reconstructs the energy of out-of-time energy deposits~\cite{TCAL-2017-01}.
Standard reconstruction methods are employed to reconstruct collision vertices from tracks in the inner detector in order to veto contributions from beam--beam collisions.
 
Jets are reconstructed using the anti-$k_{t}$ algorithm~\cite{Cacciari:2008gp,Fastjet} with a radius parameter of $R=0.4$.
Jets are required to have $\pt>20$~\GeV\ and reside within $|\eta|<4.5$.
Due to low average energy deposition in the environment studied in this search,
jet four-momentum calibration is derived from a sample of collision events with a low number of additional $pp$ interactions per BX, known as pile-up.
As such, no pile-up corrections are applied to data collected during empty BXs.
 
To reject backgrounds, reconstructed track segments in the MS are used in the signal region definitions.
Reconstructed muon segments~\cite{PERF-2015-10} are required to have at least three hits, except for CSC segments, which must have at least three $\eta$ hits and three $\phi$ hits.
For chambers in which there are two inactive layers, only two $\eta$ hits and two $\phi$ hits are necessary.
Muon segments with a failed track fit, or that have activity overlapping with adjacent events,
are rejected by including only those segments with a time-of-first-hit between $0$ and $750$~ns (where $750$~ns corresponds to the maximum drift time in the MDTs).
 
Stand-alone muons~\cite{PERF-2015-10} are used to select cosmic-ray-enriched control regions and to reject cosmic-ray-induced backgrounds in signal-enriched regions within the search sample.
Stand-alone muon tracks are formed by requiring at least two matched segments (except in the transition region between the MS barrel and endcap, where a single good-quality segment can be used).
Using this configuration, approximately 85\% of events selected in the cosmic sample have at least one stand-alone muon reconstructed.
 
 
\section{Signal models}
\label{sec:signal}

The results of this search are interpreted in the context of a split-SUSY-inspired simplified model in which gluinos are produced, forming \rhadron s,
with the gluinos decaying as $\tilde{g}\rightarrow \qqbar \tilde{\chi}_1^0$.
Signal samples were generated with gluino masses $m(\tilde{g})$ in the range $400$ to $1800$~\GeV.
Three different mass scenarios for the lightest neutralino, $\tilde{\chi}_1^0$, are considered in order to probe a range of possible kinematic signatures.
In one case the mass of the lightest neutralino $m(\tilde{\chi}_1^0)$ is fixed at $100$~\GeV,
in another more `compressed' scenario the mass difference $\deltam \equiv m(\tilde{g})-m(\tilde{\chi}_1^0)$ is fixed at $100$~\GeV, and in the third scenario $\deltam=500$~\GeV.
 
The way in which \rhadrons\ would interact with matter is not certain, but many of the uncertain aspects do not have a major impact on this analysis.
The details of the \rhadron\ mass spectrum, the formation of bound states that include gluons, the rehadronisation,
and the treatment of heavy flavour and strangeness in the initial and final states, for example, must all be treated with a specific numerical hadronic interaction model.
In this analysis the \rhadron\ simulation uses the \textit{Regge} model~\cite{deBoer:2007ii,Mackeprang:2009ad},
which makes assumptions about the mass spectra of the \rhadron\ states and their production rates.
The precise configuration used in the generation and simulation of the signal samples considered herein follows that documented in Ref.~\cite{ATL-PHYS-PUB-2019-019}.
The details of the stopping mechanisms and the probability of stopping are also discussed in this reference.
 
In the configuration detailed in Ref.~\cite{ATL-PHYS-PUB-2019-019}, a little over half of the produced \rhadrons\ are electrically neutral.
This fraction is sensitive to the probability of producing gluino--gluon bound states, since this so-called \emph{gluinoball} state is electrically neutral.
There is only a 10\% decrease in \rhadron\ stopping rate with increasing gluinoball probability over the full range of probabilities.
As such, in contrast to most searches for very long-lived particles,
the sensitivity of the analysis presented herein remains robust against variations in the gluinoball probability, being sensitive to both neutral and charged \rhadron\ states.
 
Each decay is locally isotropic as the \rhadron\ would be stationary within the detector.
The spectator partons in the final \rhadron\ do not modify the decay kinematics significantly.
Since stopped \rhadron\ decays generally occur inside the calorimeter system, the energy from the decay products tends to be deposited near the point of the decay.
Jet algorithms subsequently cluster the resulting energy depositions into one or two high-$\pt$ jets.
 
As this search is carried out using empty-BX data, signal samples were simulated without pile-up.
Any potential impact of residual pile-up on the energy measurement of signal jets is expected to be negligible,
since a pile-up jet would have to overlap geometrically with the signal jet in order to contaminate the energy reconstruction.
This assumption is supported by signal-jet energy comparisons made using signal samples simulated with and without pile-up,
where no significant difference in jet energy was observed in signal events with and without pile-up.
Even in this non-collision environment, spurious detector activity as a result of non-collision backgrounds or residual radiation in the ATLAS cavern can occur.
Such activity is not modelled in the signal simulation.
To account for this effect, simulated signal events are overlaid with events from the cavern background data sample indicated in Table~\ref{tab:datasets}.
About $92\%$ of these random events contain muon segment activity, less than $0.5\%$ contain a reconstructed muon track,
and less than $0.03\%$ contain a reconstructed jet.
 
Potential signal decays will be recorded only if the trigger accepts the event.
The sensitive time window for the trigger is estimated to be $[-10,+15]$~ns relative to the bunch crossing time, so \rhadron\ decays are placed in this time window in simulated event samples.
The detector is timed in for particles moving at the speed of light, so that particles from the collision are assigned to detector elements in the same bunch crossing.
This requires calorimeter cells further from the interaction point to have a later readout time than those closer to the interaction point.
In order to account for this effect, the decay times of the \rhadrons\ are offset by the propagation time of a particle traveling at the speed of light from the interaction point to the decay position.
 
The $\taugl$ dependence of the interpretation is evaluated using the signal \emph{live fraction} as a function of $\taugl$.
The signal live fraction defines the temporal acceptance of the trigger for a given signal scenario and depends on $\taugl$.
The live fraction for the empty-BX data considered in this analysis is shown as a function of $\taugl$ in Figure~\ref{fig:LiveFractionRun2}.
Typically, ATLAS records data continuously during periods of time in which the LHC is delivering collision data.
Each of the datasets taken while ATLAS is continuously recording is referred to as an ATLAS \textit{run}~\cite{DAPR-2018-01}.
Taking both run and bunch structure into consideration, there are two $\taugl$ regimes for the live-fraction calculation.
 
In the short lifetime ($\taugl<1$~s) regime, \rhadrons\ would usually originate, stop, and decay within a single luminosity block (LB)~\cite{DAPR-2018-01},
a data-taking interval during which detector conditions can be considered stable and that is usually about one minute in length.
For these lifetimes a simple algorithm sums the product of the probability for \rhadrons\ to be produced in each BCID within the LB and the probability for them
to decay in a subsequent empty BCID.
Per-BCID instantaneous luminosities are included in this calculation, with the assumption that the instantaneous luminosity for a given BCID is constant within an LB.
This probability, multiplied by the trigger-only live fraction,
summed and weighted by integrated luminosity over all the LBs in a data-taking period, is the live fraction for a short lifetime gluino.
The live fraction becomes significant for gluino lifetimes of $\taugl \sim 100$~ns, corresponding to the minimum time between an empty BX considered in this analysis and the previous colliding BX.
The turn-on is faster for 2018 data, where the inclusion of empty BXs closer to the last filled BX boosts sensitivity in this small $\taugl$ regime.
 
As $\taugl$ approaches a full orbit of the LHC (${\sim}100~\mu$s), the precise structure of the bunches is no longer relevant,
and the signal live fraction is dominated by the fraction of empty BCIDs in the LHC ring and the trigger live fraction.
Here the trigger-only live fraction is defined as the fraction of the total physics data-taking time corresponding to the trigger live time.
At this point the signal live fraction flattens out, reaching a plateau at the fraction of empty bunches in the machine multiplied by the trigger-only live fraction.
This plateau marks the maximum signal live fraction across the $\taugl$ range probed by this search, and corresponds to 18\% in 2017 and 16\% in 2018.
The maximum live fraction is higher in 2017 due to there being a larger number of empty BXs in the LHC ring during 2017 data-taking.
 
For longer lifetimes of the order of a minute, gluinos being produced in one LB and decaying in a subsequent LB is an important effect.
The signal live fraction becomes a weighted sum of the probability of the decay being captured in the live time of a subsequent LB over all possible production LBs in a given run.
For lifetimes approaching 1~hour, most gluinos will decay either during the same run, or otherwise before the next run begins,
with the signal live fraction decreasing as a result of the latter cases.
For even longer lifetimes, the number of potentially stopped gluinos present in the detector at any given time depends on the run structure,
and the integrated luminosity delivered during previous ATLAS runs becomes more important in accounting for potential gluino production in earlier runs.
This search considers the potential production of gluinos during the 2017 and 2018 data-taking periods only, as including the decays of gluinos produced in preceding years would only have an effect for lifetimes above $10^9$~s, where this search has no significant sensitivity.
For these lifetimes the impact on analysis sensitivity is negligible due to the exponential nature of the particle lifetime distribution,
and consequently small acceptance in the available 2017--2018 live time for potential signal production pre 2017.

\begin{figure}
\centering
\includegraphics[width=0.7\textwidth]{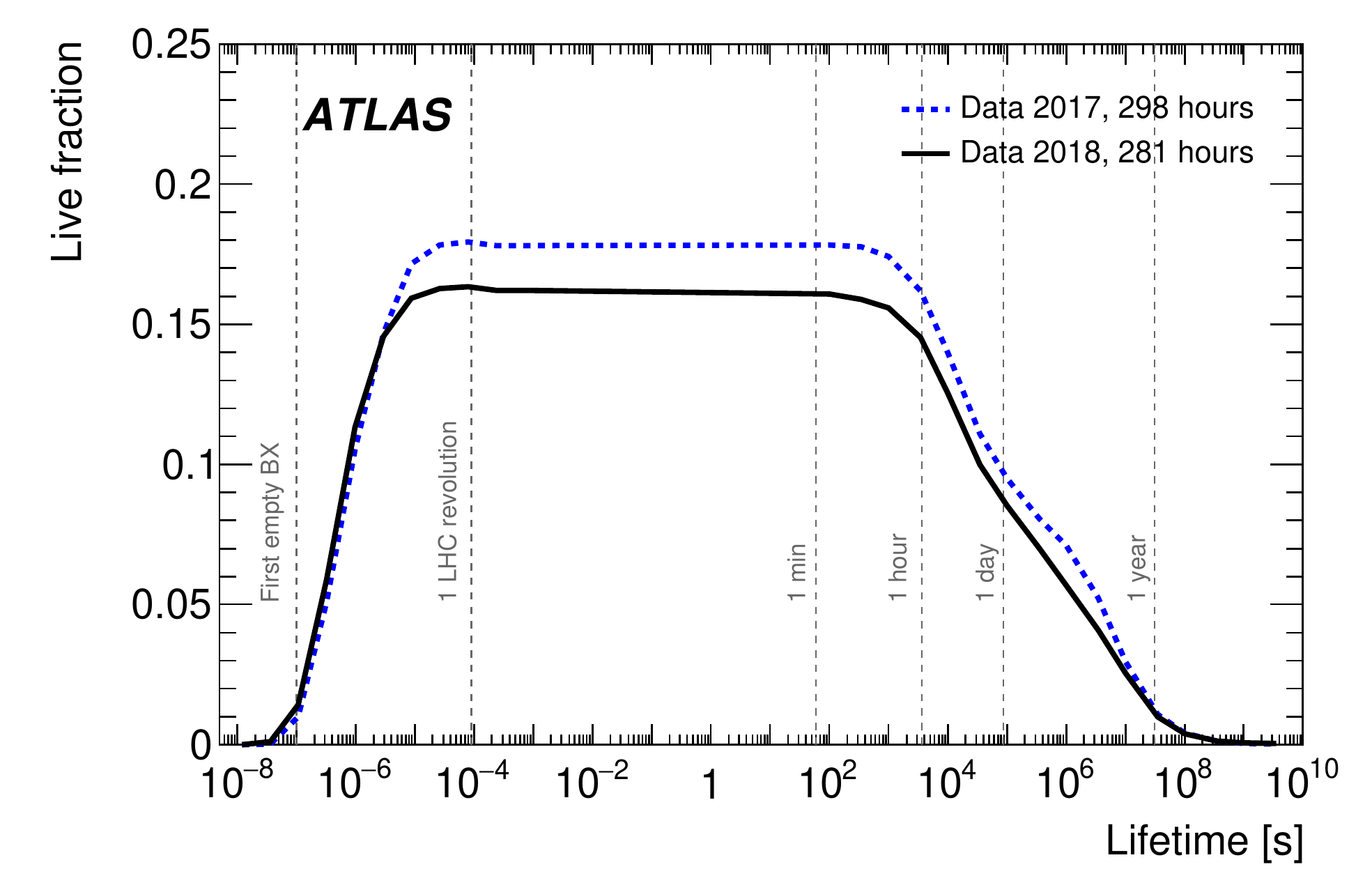}
\caption{The live fraction for empty-BX data considered in this analysis as a function of particle mean proper lifetime for data collected during 2017 and 2018. For small lifetimes, per-BCID luminosities and the LHC bunch structure are used to calculate the live fraction until it flattens at ${\sim}100~\mu$s, where the live-time is dominated by the trigger live fraction and the bunch structure. At large lifetimes, the run schedule throughout the year dominates the behavior.}
\label{fig:LiveFractionRun2}
\end{figure}
 
The expected number of signal events falling in the signal regions (SRs) for this search can be calculated as
 
\begin{equation*}
N^{\mathrm{SR}}_{\text{events}} = L^{\text{int}} \times \sigma_{\tilde{g}\tilde{g}} \times 2 \times \epsilon^{\mathrm{SR}} \times f_{\text{stopping}} \times (\text{live fraction})
\end{equation*}
 
\noindent where $L^{\text{int}}$ corresponds to the total integrated luminosity delivered, $\sigma_{\tilde{g}\tilde{g}}$ is the gluino production cross-section and
$\epsilon^{\mathrm{SR}}$ is the signal selection efficiency for the signal regions.
For a signal model with $m(\tilde{g},\tilde{\chi}^0_1)=(1400,100)$~\GeV, $\epsilon^{\mathrm{SR}}$ ranges between 25\% and 34\% for the signal regions considered in this search.
The stopping fraction $f_{\text{stopping}}$ corresponds to the fraction of produced \rhadrons\ that are expected to come to a stop within the detector.
For gluino \rhadrons\ this fraction varies from 4\% to 8\% for gluino masses between $0.6$ and $2.6$~\TeV,
with the stopping fraction increasing as a function of gluino mass.
There is a factor of two to account for the fact that the gluinos are pair-produced.
The gluino decays are unlikely to happen simultaneously,
hence there are potentially two chances to detect the decay per produced gluino pair.
The signal models used in the interpretation of the search results are normalised following this procedure.
 
\FloatBarrier
 
\section{Event selection}
\label{sec:selection}
 
Events used in this analysis must pass the following preselection requirements.
In each event the jet with the highest \pt\ must satisfy a set of cleaning criteria to reject events triggered by detector noise or non-collision backgrounds.
Additionally, this leading jet is required to have $\pt>90$~\GeV\ and to reside within $|\eta|<2.4$ to be within the geometric acceptance of the search sample trigger.
 
To avoid selecting events where most of the energy associated with a jet could be produced by localised noise,
events are rejected if the leading jet has ${>}90$\% of its energy associated with a single constituent cluster or layer within the LAr calorimeter.
Potential background from noise bursts~\cite{LARG-2013-01} in the LAr calorimeter is rejected via a veto on events where the leading jet's largest energy deposit is located in
the EM endcap calorimeter, where noise bursts originate.
Noise-induced jets in the HEC are removed by imposing the HEC-specific cleaning of the \textit{BadLoose} selection criteria described in Ref.~\cite{ATLAS-CONF-2015-029}. These cleaning requirements reject approximately 20\% of events in the signal samples used in this analysis.
 
To effectively remove collision backgrounds from the data used in this analysis, events selected in the search sample, BIB sample and cosmic sample are required to contain no primary vertex.
A primary vertex is defined as a reconstructed vertex that is associated with two or more tracks of transverse momentum $\pT > 500$~\MeV.
In empty BXs in 2018 (2017), less than 0.0005\% (0.00002\%) of events passing the search sample trigger and standard data-quality requirements were found to contain a primary vertex.
In contrast, over 80\% of events passing the BIB sample triggers contained a primary vertex,
which are mostly due to `ghost' collisions (where protons in the filled bunch in one beam collide with de-bunched protons -- or \textit{ghost charge} -- in the nominally empty bunch~\cite{DAPR-2012-01}).
There are no events containing a reconstructed primary vertex in the cosmic sample.

Events that enter the signal regions (SRs) are selected from the search sample, and the leading jet in the event is required to have \pt\ $>150$~\GeV.
Two SRs, each divided into three leading-jet \pt\ ranges, are defined:
 
\begin{itemize}
\item The \textit{central} SR (SRC) has an additional requirement that the leading jet reside within $|\eta|<0.8$.
These SRs are optimised for the split-SUSY-inspired scenarios targeted by this search, which tend to contain central decays.
\item The \textit{inclusive} SR (SRIncl) accepts all events in which the leading jet resides within $|\eta|<2.4$, corresponding to the acceptance dictated by the trigger.
These SRs are included to preserve acceptance for potential signal scenarios with more forward topologies~\cite{Kang:2008ea,Knapen:2017kly,Evans:2018jmd}.
\end{itemize}

Events containing a reconstructed muon are rejected from the signal regions to remove events seeded by cosmic-ray-induced jets.
Further cosmics rejection is provided by vetoing events in which the leading jet overlaps with a trajectory between two reconstructed muon segments
-- an indication that the jet is likely induced by a cosmic ray.
This is achieved via the use of the $\alpha$ variable,
which represents the minimum $\eta$--$\phi$ distance between the leading jet and a putative cosmic-ray muon (inferred from a combination of upper- and lower-hemisphere muon segments) as it passes through the calorimeter.
For cosmic-ray-induced jets, $\alpha$ takes on small values, and events are required to have $\alpha>0.2$ to enter the SRs.
In cases where $\alpha$ cannot be defined (typically $>50$\% of signal events and $8$\% of cosmic-ray events), events are accepted into the SRs.
The construction of $\alpha$ is described in detail in Section~\ref{sec:cosmics}.
 
To reject BIB the narrow BIB-jet width in the $\phi$-plane,
resulting from the relatively narrow showers expected from BIB traversing the detector parallel to the beam,
is exploited via the use of the \widthphi\ variable. This is defined as
 
\begin{equation*}
\widthphi = \frac{ \sum_{i} \pt(i) \cdot |\Delta\phi(\mathrm{jet},i)| } { \sum_{i} \pt(i) },
\end{equation*}
 
\noindent where, in the sums, $i$ runs over the constituents of the jet and $\Delta\phi(\mathrm{jet},i)$ is the distance in the $\phi$ dimension between a constituent energy cluster associated with the leading jet in the event and the reconstructed centroid of that jet.
While signal events tend to have an isotropic topology, the parallel-to-the-beam jets from BIB tend to have small \widthphi\ values,
and events entering the SRs are required to have $\widthphi>0.02$.
The characteristics and estimation of BIB are described in detail Section~\ref{sec:bib}.

The SRs are binned by leading jet $\pt$, with bin boundaries defining the ranges $150<\pt<300$~\GeV, $300<\pt<500$~\GeV\ and $\pt>500$~\GeV.
Each jet \pt\ bin offers sensitivity to signal scenarios with different \deltam.
The lowest \pt\ bin offers sensitivity to scenarios with $\deltam=100$~\GeV, while the higher jet-\pt\ bins offer coverage for scenarios with larger mass splittings.
An overview of the signal regions, separated by jet \pt\ bin boundaries, is given in Table~\ref{tab:SRs}.
In order to provide production cross-section upper limits for the signal models described in Section~\ref{sec:signal}, all bins in leading jet \pt\ are included in a profile-likelihood~\cite{Cowan:2010js} fit.
To form six discovery regions (DRs), each SR \pt-bin is made inclusive of larger leading jet \pt\.
These DRs are used individually to probe the existence of BSM physics or to assess model-independent upper limits on the number of possible signal events.
Because the sensitivity of the 2018 dataset to generic signal models is expected to dominate, the discovery regions comprise 2018 data only.
The six DRs are also summarised in Table~\ref{tab:SRs}.
 
\begin{table}[htb]
\small
\begin{center}
\caption{Overview of the signal and discovery regions. These regions are divided into two sets; one set imposes an $|\eta|<0.8$ requirement on the leading jet in the event, while the other set accepts leading jets within $|\eta|<2.4$.
Events for which $\alpha$ cannot be determined are included in the SRs and DRs.}
\vspace{0.2cm}
\resizebox{0.95\textwidth}{!}{
\begin{tabular}{lccccccc}
\hline
 
\textbf{Region}          & \textbf{Data sample} & \textbf{Number of}             & \textbf{Leading}        & \textbf{$\alpha$} & \textbf{Leading jet}  & \textbf{Leading}    \\
&                      & \textbf{muons}                 & \textbf{jet \pt\ [GeV]} &                                & \textbf{\widthphi} & \textbf{jet $|\eta|$}  \\
\hline\hline
\multicolumn{7}{l}{\textbf{Central signal region}} \\
\hline\hline
\multirow{3}{*}{SRC}         & \multirow{3}{*}{Search sample}        & \multirow{3}{*}{$0$}     & $150$--$300$~~~~    & \multirow{3}{*}{$>0.2$} & \multirow{3}{*}{$>0.02$}  & \multirow{3}{*}{$<0.8$}       \\
&                                       &                          & $300$--$500$~~~~    &                         &                           &        \\
&                                       &                          & $>500$       &                         &                           &        \\
 
\hline\hline
 
\multicolumn{7}{l}{\textbf{Inclusive signal region}} \\
\hline\hline
\multirow{3}{*}{SRIncl}      & \multirow{3}{*}{Search sample}        & \multirow{3}{*}{$0$}     & $150$--$300$~~~~     & \multirow{3}{*}{$>0.2$}& \multirow{3}{*}{$>0.02$}  & \multirow{3}{*}{$<2.4$}       \\
&                                       &                          & $300$--$500$~~~~     &                        &                           &        \\
&                                       &                          & $>500$        &                        &                           &        \\
\hline\hline
\multicolumn{7}{l}{\textbf{Central discovery regions}} \\
\hline\hline
DRC-150               & Search sample                         & \multirow{3}{*}{$0$}     & $>150$    & \multirow{3}{*}{$>0.2$} & \multirow{3}{*}{$>0.02$}  & \multirow{3}{*}{$<0.8$}       \\
DRC-300               & (2018 data only)                      &                          & $>300$    &                         &                           &        \\
DRC-500               &                                       &                          & $>500$       &                         &                           &        \\
 
\hline\hline
 
\multicolumn{7}{l}{\textbf{Inclusive discovery regions}} \\
\hline\hline
DRIncl-150            & Search sample                         & \multirow{3}{*}{$0$}     & $>150$     & \multirow{3}{*}{$>0.2$}& \multirow{3}{*}{$>0.02$}  & \multirow{3}{*}{$<2.4$}       \\
DRIncl-300            & (2018 data only)                      &                          & $>300$     &                        &                           &        \\
DRIncl-500            &                                       &                          & $>500$        &                        &                           &        \\
\hline
\end{tabular}
}
\label{tab:SRs}
\end{center}
\end{table}

\FloatBarrier
 
\section{Background estimation}
\label{sec:bg}
 
The main backgrounds considered in this analysis are as follows:
\begin{itemize}
\item \textbf{Cosmic rays:}
Muons from cosmic rays can induce energetic jets in the calorimeter while traversing the detector.
The methods used to minimise this background and predict the remaining contribution to the SRs are discussed in Section~\ref{sec:cosmics}.
\item \textbf{Beam-induced backgrounds:}
Beam protons can interact with upstream collimators, residual gas within the beam pipe, or the beam pipe itself, resulting in energetic muons traversing the detector in the horizontal plane.
The methods used to reduce and estimate the extent of these backgrounds are detailed in Section~\ref{sec:bib}.
\end{itemize}

In both the cosmic-ray- and beam-induced backgrounds, the \pt\ distribution of the leading jet falls steeply,
and the background estimation strategy relies on the modelling of the shapes of these distributions.
Jet \pt\ templates are extracted for both cosmics and BIB in dedicated control regions (CRs).
The background estimation strategy and validation is described in detail in Sections~\ref{sec:cosmics} and~\ref{sec:bib},
but the control and validation regions are outlined below.
 
To model the cosmics background in the SRs, a `cosmic-tag' selection is applied to data collected in both the search and the BIB samples to construct cosmics-enriched control regions.
The BIB and search samples are combined to provide the highest possible number of events in the template construction.
A cosmic-tag is enforced by the requirement that events contain at least one central muon ($|\eta|<1.4$) and have $\alpha<0.2$.
Although the cosmic sample provides a pure sample of cosmic-ray-induced events,
the sample size is not sufficiently large to extract a statistically robust jet-\pt\ template.
Instead, a cosmics CR is defined for each of the SRs, with a selection otherwise matching that of the SR as closely as possible.
These CRs are denoted CRC-cos, corresponding to SRC, and CRIncl-cos, corresponding to SRIncl.
The shape of the jet \pt\ distribution is taken from each cosmics CR and is extrapolated to the SR of interest via a transfer factor obtained from a cosmic sample,
as described in detail in Section~\ref{sec:cosmics}.
The level of signal contamination in the 2016 cosmic sample from potential \rhadron\ production in earlier $pp$ collision data-taking was found to be negligible.
 
For BIB, the jet \pt\ template is derived from the BIB sample.
BIB CRs are constructed with a selection closely matching that of SRC (to form CRC-bib) and SRCIncl (to form CRIncl-bib), but with the \widthphi\ requirement lowered to $\widthphi>0.01$ to accept more background events.
The level of signal contamination in the BIB CRs for a 1~\TeV\ gluino is found to be negligible.
The BIB-jet \pt\ templates are then normalised in dedicated normalisation regions (NRs) using search sample data.
Each normalisation region has a selection identical to that of the SR of interest, but requires a leading jet \pt\ between $90$ and $150$~\GeV.
The control and normalisation region definitions are given in Table~\ref{tab:regions}.
 
\begin{table}[htb]
\small
\begin{center}
\caption{Overview of the control and normalisation regions used to estimate the contribution of background processes to the SRs.
The main requirements that distinguish the control and normalisation regions from the signal regions are indicated in boldface.
Events for which $\alpha$ cannot be determined are included in the $\alpha>0.2$ selection CR(Incl/C)-bib and NR(Incl/C)-bib.
}
\vspace{0.2cm}
\resizebox{0.95\textwidth}{!}{
\begin{tabular}{lccccccc}
\hline
 
\textbf{Region}          & \textbf{Data sample} & \textbf{Number of}             & \textbf{Leading}        & \textbf{$\alpha$} & \textbf{Leading jet}  & \textbf{Leading}    \\
&                      & \textbf{muons}                 & \textbf{jet \pt\ [GeV]} &                                & \textbf{\widthphi} & \textbf{jet $|\eta|$}  \\
\hline\hline
\multicolumn{7}{l}{\textbf{Central control and normalisation regions}} \\
\hline\hline
CRC-cos                  & Search sample \& BIB sample           & \textbf{$\ge 1$ ($|\eta|<1.4$)}   & $>90$        & \textbf{$<0.2$}                        & $>0.02$              & \multirow{3}{*}{$<0.8$}       \\
CRC-bib                  & \textbf{BIB sample}                            & $0$                      & $>90$        & $>0.2$                        & $>0.01$              &        \\
NRC-bib                  & Search sample                         & $0$                      & \textbf{$90$--$150$}~~~~     & $>0.2$                        & $>0.02$              &        \\
\hline\hline
 
\multicolumn{7}{l}{\textbf{Inclusive control and normalisation regions}} \\
\hline\hline
CRIncl-cos                  & Search sample \& BIB sample           & \textbf{$\ge 1$ ($|\eta|<1.4$)}   & $>90$        & \textbf{$<0.2$}                        & $>0.02$            & \multirow{3}{*}{$<2.4$}       \\
CRIncl-bib                  & \textbf{BIB sample}                   & $0$                      & $>90$        & $>0.2$                        & $>0.01$            &        \\
NRIncl-bib                  & Search sample                         & $0$                      & \textbf{$90$--$150$}~~~~     & $>0.2$                        & $>0.02$            &        \\
\hline
\end{tabular}
}
\label{tab:regions}
\end{center}
\end{table}

\subsection{Cosmic-ray muon background}
\label{sec:cosmics}
 
Events containing cosmic-ray muons can enter into the event selection of this analysis when a traversing muon induces an energetic jet in the calorimeter.
This background is dominated by the emission of a highly energetic bremsstrahlung photon from a high-momentum muon.
The \pt\ distribution of cosmic-ray-induced jets has a longer high-\pt\ tail than the corresponding distribution of jets from beam-induced backgrounds,
making it the dominant background process in the highest jet-\pt\ range targeted by this search.
 
As a cosmic-ray muon traverses the detector from top to bottom it passes through the MS.
The presence of reconstructed muon segments in opposite hemispheres of the detector can be an indication of a cosmic-ray muon event,
although energetic jets produced in gluino decays could punch through from the calorimeter into the MS, also resulting in reconstructed segments.
Spurious muon segments can also occur independently of potential signal decays as a result of cavern background,
contributing muon segment activity to potential signal events that is uncorrelated with the signal decays.
This makes relying on muon segment information alone impractical, as it results in decreased signal selection efficiency.
The $\alpha$ variable uses the relative geometric location of a pair of reconstructed muon segments combined with the location of the leading jet in the event
(assumed to be seeded by the same traversing cosmic-ray muon as the muon segments).
This concept is illustrated in Figure~\ref{fig:alphacos_cartoon}.
Here the left and middle panels show a jet positioned within the calorimeter, near the path of a traversing cosmic-ray muon.
The right panel shows how the $\alpha$ variable is defined, based on geometric information associated with these objects.

\begin{figure}
\centering
\includegraphics[width=\textwidth]{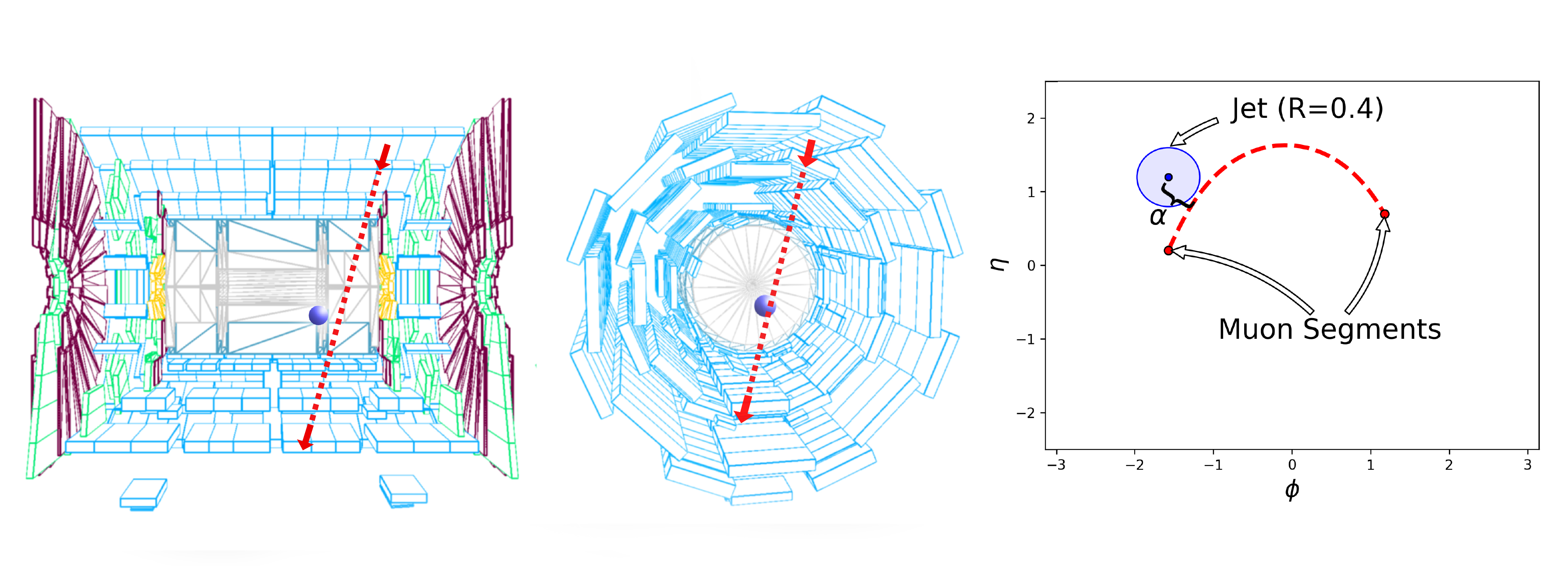}
\caption{Example illustration of the construction of the $\alpha$ variable. Left: Longitudinal view demonstrating the assumed cosmic-ray muon trajectory for a given pair of muon segments and a leading jet. The location of the calorimeter energy deposition forming this jet is shown as a blue sphere.
Middle: Transverse view demonstrating the assumed cosmic-ray muon trajectory for a given pair of muon segments and a leading jet. Right: An $\eta$--$\phi$ plane view of the assumed cosmic-ray muon trajectory between a pair of muon segments, and the $\Delta R$ that corresponds to the $\alpha$ variable in this example. In this case the $\alpha$ value is greater than the size of the jet radius parameter ($R=0.4$).}
\label{fig:alphacos_cartoon}
\end{figure}

To calculate $\alpha$, reconstructed muon segments in an event are grouped into pairs.
A pair is formed if the trajectory measured in at least one of the two muon segments points in the direction of the other,
with each pair containing one upper- and one lower-hemisphere muon segment.
In order to recover inefficiencies due to the MS gap at $\eta=0$,
the direction vector of each segment is also extrapolated to the opposite side of the detector in $\phi$ to determine whether it crosses the $\eta=0$ plane within the radial range otherwise instrumented by the MS.
Should this be the case, a `dummy' segment is inserted at that position and used in the calculation of $\alpha$,
to acknowledge that a traversing cosmic-ray muon could escape through the gap without seeding a muon segment.
 
For each pair of muon segments in an event,
a $\Delta R$ value is calculated for the pair of points where the axis of the leading jet and the 3D straight line connecting the muon segment pair are closest to each other.
The $\alpha$ variable is defined as the minimum of these `points of closest approach' $\Delta R$ values in the event.
This is illustrated in the right panel of Figure~\ref{fig:alphacos_cartoon} in the $\eta$--$\phi$ plane.
The points of closest approach must correspond to a location consistent with being within the volume of the calorimeter.
If this is not the case, or if no suitable pair of muon segments is identified, $\alpha$ is undefined and the event is not rejected.
This occurs in 8\% of the events in the cosmic sample, 56\% of the events in the signal model with $\deltam=1300$~\GeV, and 78\% of the events in the signal model with $\deltam=100$~\GeV.
For events where the leading jet is induced by a cosmic-ray muon, $\alpha$ takes on small values, as shown in Figure~\ref{fig:alphaCos},
and events are rejected from the SRs if they satisfy $\alpha<0.2$. For signal events, a value of $\alpha$ may be defined due to the leakage of hadronic activity out of the calorimeter and the cavern background effects described above.

\begin{figure}
\centering
\includegraphics[width=0.7\textwidth]{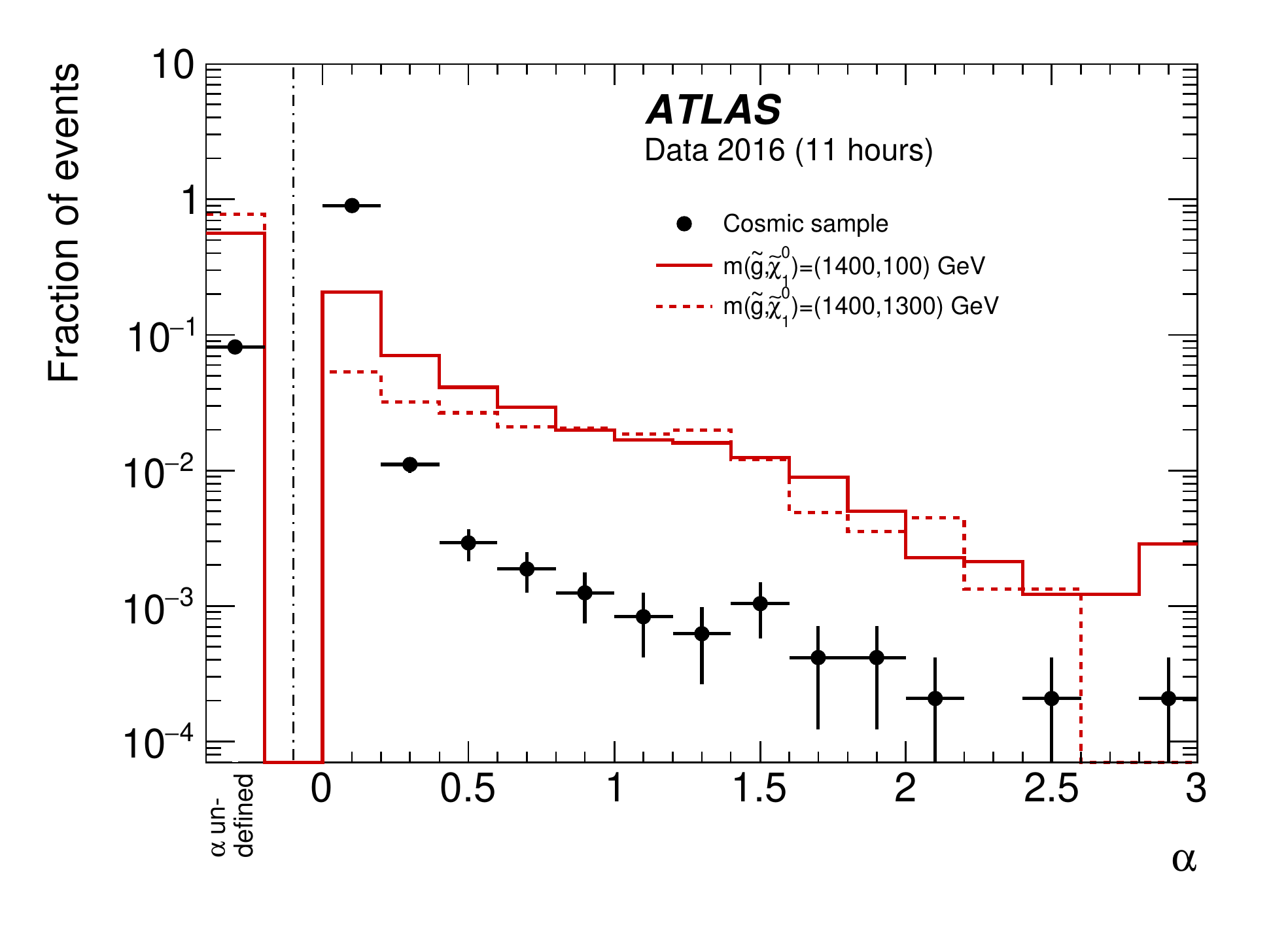}
\caption{The $\alpha$ distribution is shown for events passing preselection requirements in the cosmic sample.
Example signal models with a gluino mass of $1.4$~\TeV\ and mass differences $\deltam=100$~\GeV\ and 1300~\GeV\ are overlaid.
In cases where all eligible muon segments reside in one hemisphere of the detector, or where no segments have been reconstructed in the event, $\alpha$ cannot be calculated.
The first bin represents the events for which $\alpha$ cannot be calculated because the imposed requirements are not met.
The last bin contains the overflow.
}
\label{fig:alphaCos}
\end{figure}

To estimate the remaining cosmics background expected in the SRs,
\pt\ templates for cosmic-ray-induced jets are extracted from the cosmics CRs (CRIncl-cos and CRC-cos).
The selection applied in these CRs provides a highly pure sample of cosmic-ray events in data,
with negligible contamination from BIB processes and potential signal decays.
The shape of the resulting jet \pt\ template is validated by comparison with the same distribution taken directly from the cosmic sample.
The shapes of the distributions are consistent within statistical uncertainties.
 
The cosmic sample is used to derive a transfer factor between the control and signal regions.
The same transfer factor is used to extrapolate between the CR and SR for both the inclusive and central SRs,
as the differing $\eta$ requirements have no significant impact on the transfer factor.
The transfer factor is derived from the $|\eta|<2.4$ selection as follows:
\begin{equation*}
F_\mathrm{transfer} = \frac{N_{\text{events (SRIncl)}}^{\text{cosmic sample}}}{N_{\text{events (CRIncl-cos)}}^{\text{cosmic sample}}}
\end{equation*}
 
\noindent where $N_{\text{events (SRIncl)}}^{\text{cosmic sample}}$ and $N_{\text{events (CRIncl-cos)}}^{\text{cosmic sample}}$ correspond to the total number of events passing SR- and CR-like selections, respectively, in the cosmic sample.
The \pt\ template for cosmic-ray-induced jets is first normalised to the total number of events entering the CR from the search sample.
A transfer factor of $0.018 \pm 0.006$ is then applied to the leading-jet \pt\ template extracted from the corresponding cosmics CR.
To account for differences in the shape of the jet \pt\ distribution when extrapolating from $\alpha<0.2$ to $\alpha>0.2$,
a loose selection is applied to the cosmic sample used to derive per-jet \pt\ bin weights that can be used to reweight the jet \pt\ template taken from the cosmics CRs when extrapolating to the SRs.
The impact of this reweighting is small at low \pt, with reweighting factors reaching 1.6 in the highest jet-\pt\ bin.
The normalisation of the template remains constant upon applying these weights, so as to preserve the total expected event counts in the SR.
Following this procedure, the expected contribution from cosmics in the relevant SR is obtained.

\FloatBarrier
\subsection{Beam-induced background}
\label{sec:bib}
 
The primary component of BIB is beam halo, which occurs as a result of beam protons interacting with
upstream collimators, residual gas inside the beam pipe, or the beam pipe itself, resulting in energetic muons traversing the detector in the horizontal plane.
The contribution from beam--gas interactions inside the length of beam pipe within the detector volume is particularly small in empty BXs,
and these events are further reduced by the requirement that there be no reconstructed vertex in selected events.
In order for beam halo to contribute to this search,
the interactions must take place sufficiently far upstream from the IP to allow such muons to reach relatively large radii in the ATLAS detector, reaching the calorimeters and seeding energetic jets.
Owing to bending via the dipole magnets of the LHC, beam-halo muons tend to arrive in ATLAS in the horizontal plane~\cite{DAPR-2014-01},
with the resulting jets being of larger width in $\eta$ than in $\phi$.
The width of BIB-jets in the $\phi$ direction tends to be small,
due to the trajectory of the traversing muons being approximately constant in $\phi$ and the fact that the resulting energy deposition is primarily electromagnetic (yielding narrow showers).
This is illustrated in Figure~\ref{fig:bib-widthPhi}.
A requirement of $\widthphi>0.02$ is applied to reject BIB in the SRs.
 
\begin{figure}
\centering
\includegraphics[width=0.7\textwidth]{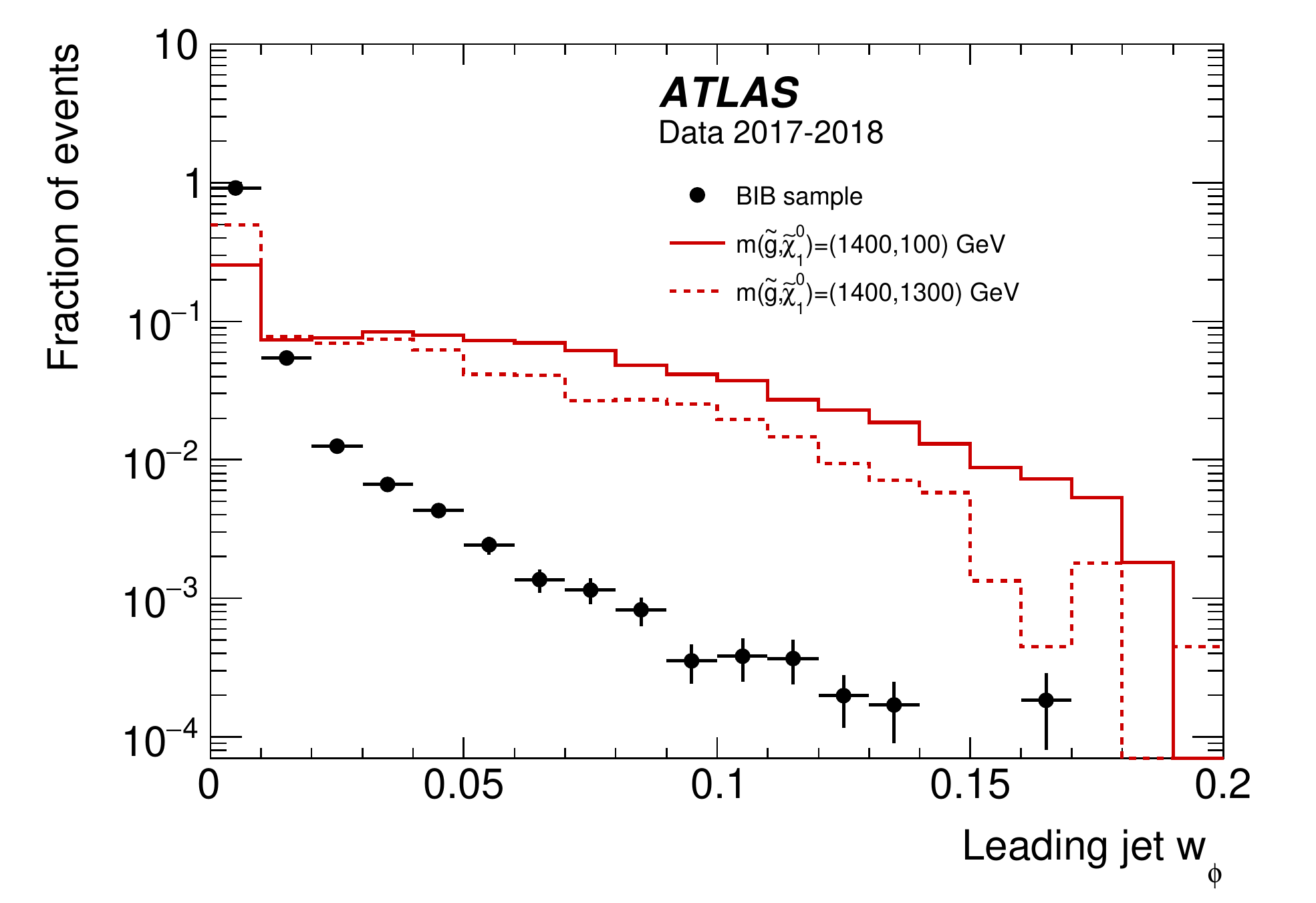}
\caption{Leading-jet \widthphi\ distribution for events passing preselection requirements in the BIB sample.
Example signal models with a gluino mass of $1.4$~\TeV\ and a mass difference $\deltam=100$~\GeV\ and 1300~\GeV\ are overlaid.
The last bin contains the overflow.}
\label{fig:bib-widthPhi}
\end{figure}

Events from unpaired BXs, where only one of the two counter-rotating beams has a filled bunch,
provide a BIB-enriched data sample in which the contribution from cosmic rays is small compared to that in empty-BX data.
The relative intensity of BIB as a function of time within the $25$~ns BCID time window differs between unpaired- and empty-BX data.
In unpaired-BX data the dominant BIB contribution comes from BIB that is in-time with the `nominal' RF bucket,
i.e.\ the $2.5$~ns bucket within the filled BCID that contains the proton bunch.
In contrast, BIB in empty-BX data tends to be evenly distributed across all RF buckets,
although the production mechanism remains the same.
A dedicated systematic uncertainty is applied to the BIB prediction to account for differences in BIB characteristics between empty- and unpaired-BX data,
as described in Section~\ref{sec:syst}.
 
To estimate the contribution from BIB to the SRs, a leading-jet \pt\ template is taken from the corresponding BIB CR, which is constructed using the BIB sample.
To facilitate this, the leading-jet \widthphi\ requirement is loosened to ${>}0.01$,
since the SR $\widthphi>0.02$ cut significantly reduces the size of the BIB sample.
The BIB template obtained in this region is consistent with the distribution obtained using a tighter $\widthphi>0.02$ selection within statistical uncertainties.
In order to remove contamination from cosmic-ray-induced events the cosmics background contribution must be estimated and removed from the BIB CRs.
This is done by taking the cosmics template from a cosmics CR-like region, but with a lowered \widthphi\ threshold of $0.01$,
and following the steps outlined in Section~\ref{sec:cosmics} to estimate the cosmics background in the corresponding BIB CR.
The resulting cosmics template is then subtracted from the BIB template to yield a BIB-only jet-\pt\ template,
which is then normalised in the BIB NR corresponding to the SR of interest.
The cosmics contribution to this normalisation region is first subtracted using the 90--150~\GeV\ jet-\pt\ bins of the SR cosmics template, derived as described in Section~\ref{sec:cosmics}.
Since BIB rates and characteristics depend on machine and beam conditions, and LHC operating conditions differed between 2017 and 2018, this procedure is performed separately for 2017 and 2018 data.
The BIB normalisation factors for 2017 data are $0.70 \pm 0.06$ (SRIncl) and $0.65 \pm 0.10$ (SRC), while for 2018 data they are $0.55 \pm 0.03$ (SRIncl) and $0.61 \pm 0.06$ (SRC).
 
\FloatBarrier
\subsection{Background estimate validation}
\label{sec:VRs}
 
The background estimation procedures outlined in Sections~\ref{sec:cosmics} and ~\ref{sec:bib} are tested using dedicated validation regions (VRs).
A summary of the kinematic selections used in the validation regions is given in Table~\ref{tab:validation-regions}.
In each of these validation regions, the BIB templates are normalised in the 90--150~\GeV\ leading-jet \pt\ range.
The BIB normalisation regions used for each VR are also summarised in Table~\ref{tab:validation-regions}.
Validation regions are defined for both the leading-jet $\eta$-central and $\eta$-inclusive event selections,
and the background estimates are validated separately for 2017 and 2018 data, owing to the different BIB characteristics year-to-year.
For brevity, the 2017 and 2018 VRs are combined in the following description.
 
The $\alpha$ VRs (VRIncl-$\alpha$ and VRC-$\alpha$) are designed to test the extrapolation and reweighting of the cosmics \pt\ template from the cosmics CRs to $\alpha>0.2$.
The central-muon requirement ensures a region dominated by cosmic-ray-induced events in order to test this.
The BIB VRs (VRIncl-bib and VRC-bib) have an $\alpha$ requirement identical to that of the SR,
but a lower \widthphi\ requirement in order to obtain a data sample dominated by BIB events to validate the BIB modelling at high $\alpha$.
The \widthphi\ VRs (VRIncl-\widthphi\ and VRC-\widthphi) are constructed with the same leading-jet \widthphi\ requirement as the SR,
but with the $\alpha$ requirement inverted.
This results in a region where the modelling of both the cosmics and the BIB can be checked at high \widthphi.
In this region, no reweighting is applied to the cosmics template, as there is no extrapolation over $\alpha$.
For VRIncl-$\alpha$ and VRIncl-\widthphi\ (VRC-$\alpha$ and VRC-\widthphi) the cosmics template is extracted from CRIncl-cos (CRC-cos).
For VRIncl-bib (VRC-bib) the \widthphi\ selection of CRIncl-cos (CRC-cos) is modified to match that of the BIB VRs ($0.01<\widthphi<0.02$).
The BIB templates for VRIncl-$\alpha$  and VRIncl-bib (VRC-$\alpha$ and VRC-bib) are taken from CRIncl-bib (CRC-bib),
but for VRIncl-\widthphi\ (VRC-\widthphi) the CRIncl-bib (CRC-bib) $\alpha$ requirement is inverted to match that of the \widthphi\ VRs ($\alpha<0.2$).
For a 1~\TeV\ gluino the level of signal contamination in the VRs is ${<}10$\%.
The exceptions to this are the single bins above $500$~\GeV\  in VR-$\alpha$ (${<}20$\% for all gluino masses) and VR-bib.
In the bin above $500$~\GeV\ in VR-bib, potential signal contamination is ${<}25$\% for models with a 1.4~\TeV\ gluino and a 100~\GeV\ neutralino,
with the signal population decreasing significantly for 1.4~\TeV\ gluino models with smaller mass-splittings.
The signal contamination can be up to $100$\% for models with lower gluino masses and large mass-splittings.

\begin{table}[htb]
\caption{Overview of the validation regions and the low-\pt\ regions used to normalise the BIB templates for the corresponding validation region.
A central set and an inclusive set of these regions are defined to test the background modelling in both SRC and SRIncl.
The VRs are also checked separately using data from 2017 and from 2018 to ensure consistent background modelling in both years.
The main requirements that distinguish the validation regions from the signal regions, and the normalisation regions from the validation regions, are indicated in boldface.
Events for which $\alpha$ cannot be determined are included in the $\alpha>0.2$ selection in VR(Incl/C)-$\alpha$, VR(Incl/C)-bib and the corresponding NRs.
}
\small
\begin{center}
\resizebox{0.95\textwidth}{!}{
\begin{tabular}{lcccccc}
\hline
 
\textbf{Region}         & \textbf{Data} & \textbf{Number of}             & \textbf{Leading}        & \textbf{$\alpha$} & \textbf{Leading jet}              & \textbf{Leading}\\
& \textbf{sample}       & \textbf{muons}                 & \textbf{jet \pt\ [GeV]} &                        & \textbf{\widthphi}  & \textbf{jet $|\eta|$}\\
\hline\hline
\multicolumn{7}{l}{\textbf{Central validation regions}} \\
\hline\hline
VRC-$\alpha$    & \multirow{3}{*}{Search sample} & \textbf{$\ge 1$ ($|\eta|<1.4$)}   & \multirow{3}{*}{$>150$} & $>0.2$          & ~~$>0.02$       &  \multirow{3}{*}{$<0.8$}   \\
VRC-bib                      &                                & $0$                      &                         & $>0.2$          & \textbf{$0.01$--$0.02$}~~~   &         \\
VRC-\widthphi  &                                & $0$                      &                         & \textbf{$<0.2$}          & ~~$>0.02$       &         \\
\hline\hline
\multicolumn{7}{l}{\textbf{Inclusive validation regions}} \\
\hline\hline
VRIncl-$\alpha$    & \multirow{3}{*}{Search sample} & \textbf{$\ge 1$ ($|\eta|<1.4$)}  & \multirow{3}{*}{$>150$} & $>0.2$          & ~~$>0.02$       &  \multirow{3}{*}{$<2.4$}   \\
VRIncl-bib                      &                                & $0$                      &                         & $>0.2$          & \textbf{$0.01$--$0.02$}~~~   &         \\
VRIncl-\widthphi  &                                & $0$                      &                         & \textbf{$<0.2$}          & ~~$>0.02$       &         \\
\hline\hline
\multicolumn{7}{l}{\textbf{Central BIB normalisation regions}} \\
\hline\hline
NRC-$\alpha$  & \multirow{3}{*}{Search sample} & $\ge 1$ ($|\eta|<1.4$)   & \multirow{3}{*}{\textbf{$90$--$150$}} & $>0.2$            & ~~$>0.02$       &   \multirow{3}{*}{$<0.8$}   \\
NRC-bib                  &                                & $0$                      &                           & $>0.2$            & $0.01$--$0.02$~~~   &     \\
NRC-\widthphi    &                          & $0$                      &                           & $<0.2$            & ~~$>0.02$       &     \\
\hline\hline
\multicolumn{7}{l}{\textbf{Inclusive BIB normalisation regions}} \\
\hline\hline
NRIncl-$\alpha$  & \multirow{3}{*}{Search sample} & $\ge 1$ ($|\eta|<1.4$)   & \multirow{3}{*}{\textbf{$90$--$150$}} & $>0.2$            & ~~$>0.02$       &   \multirow{3}{*}{$<2.4$}   \\
NRIncl-bib                  &                                & $0$                      &                           & $>0.2$            & $0.01$--$0.02$~~~   &     \\
NRIncl-\widthphi    &                          & $0$                      &                           & $<0.2$            & ~~$>0.02$       &     \\
\hline
\end{tabular}
}
\label{tab:validation-regions}
\end{center}
\end{table}
 
The resulting background predictions are compared with the observed data in each of the VRs in Figure~\ref{fig:VRs}.
The predicted distributions are in good agreement with the observed data in the VRs.
 
\begin{figure}
\centering
\includegraphics[width=0.44\textwidth]{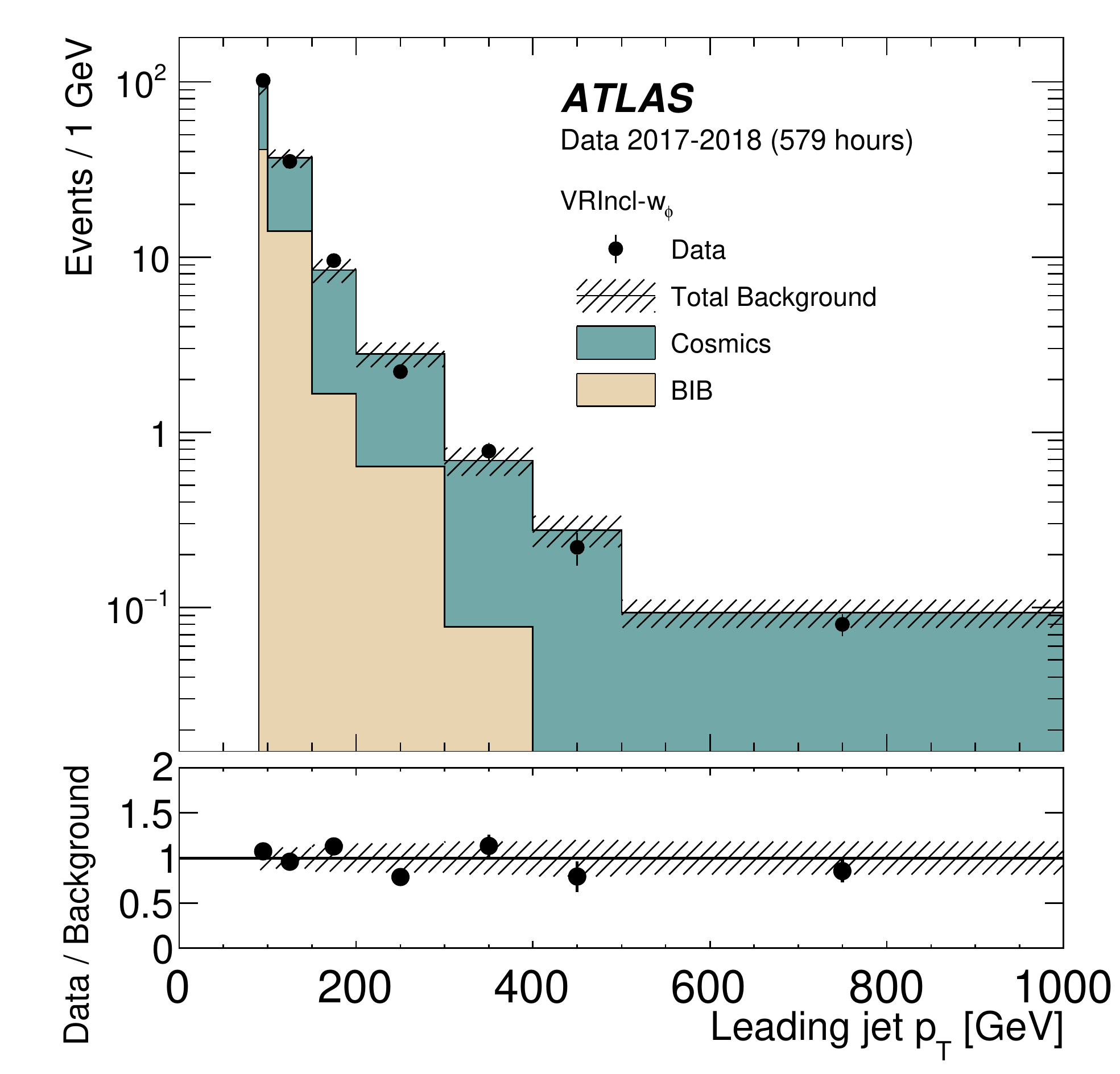}
\includegraphics[width=0.44\textwidth]{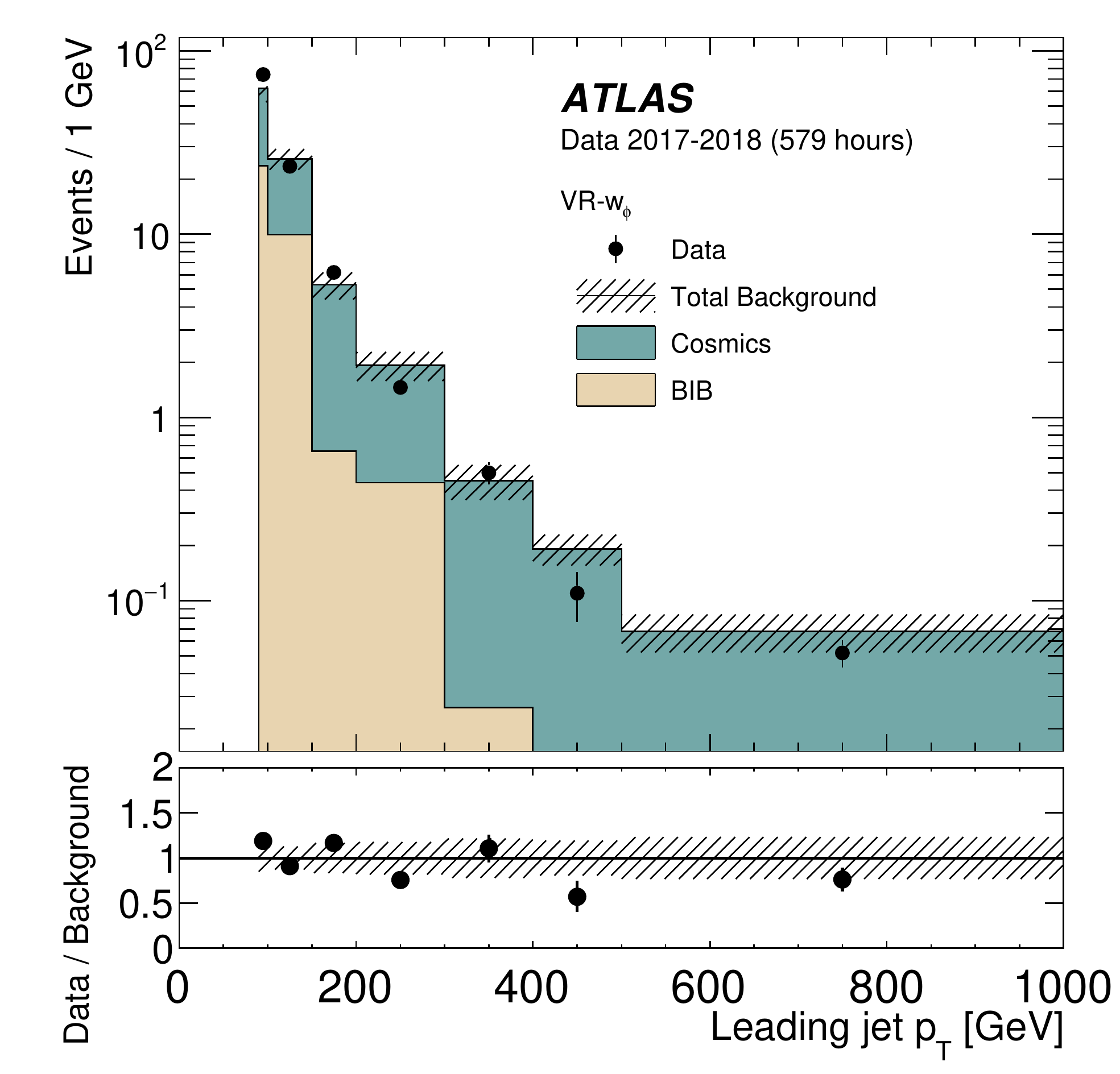}
\includegraphics[width=0.44\textwidth]{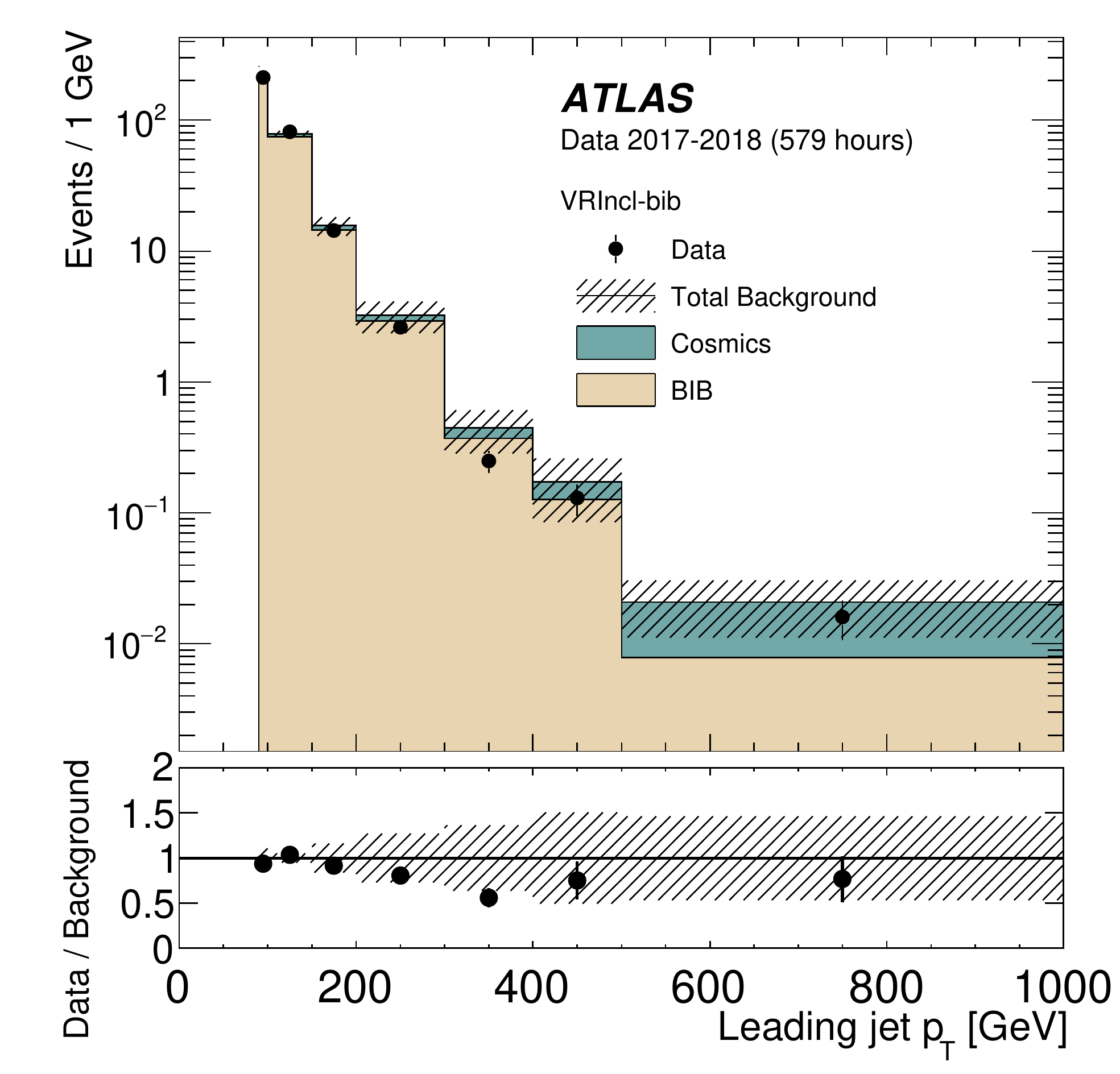}
\includegraphics[width=0.44\textwidth]{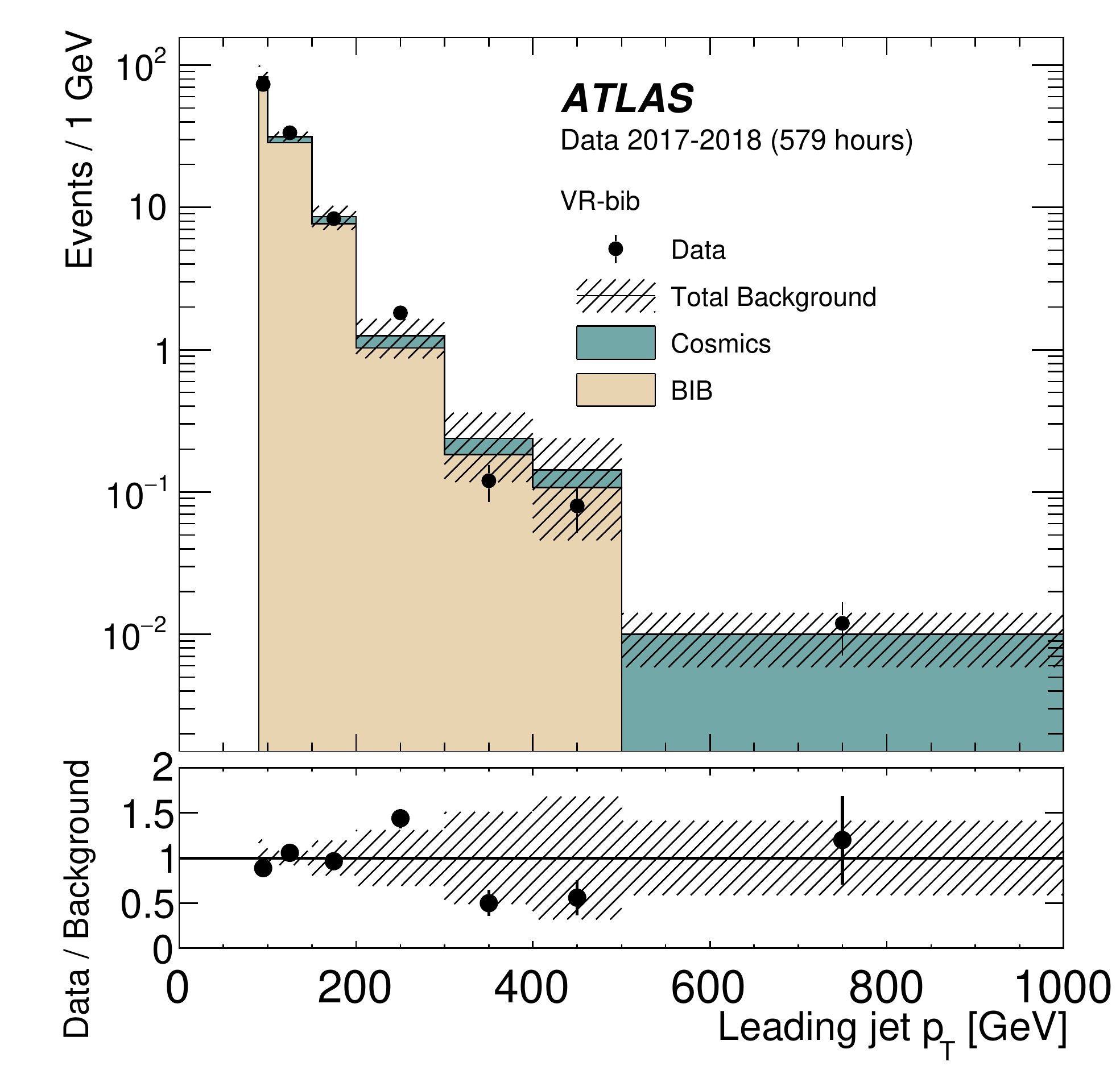}
\includegraphics[width=0.44\textwidth]{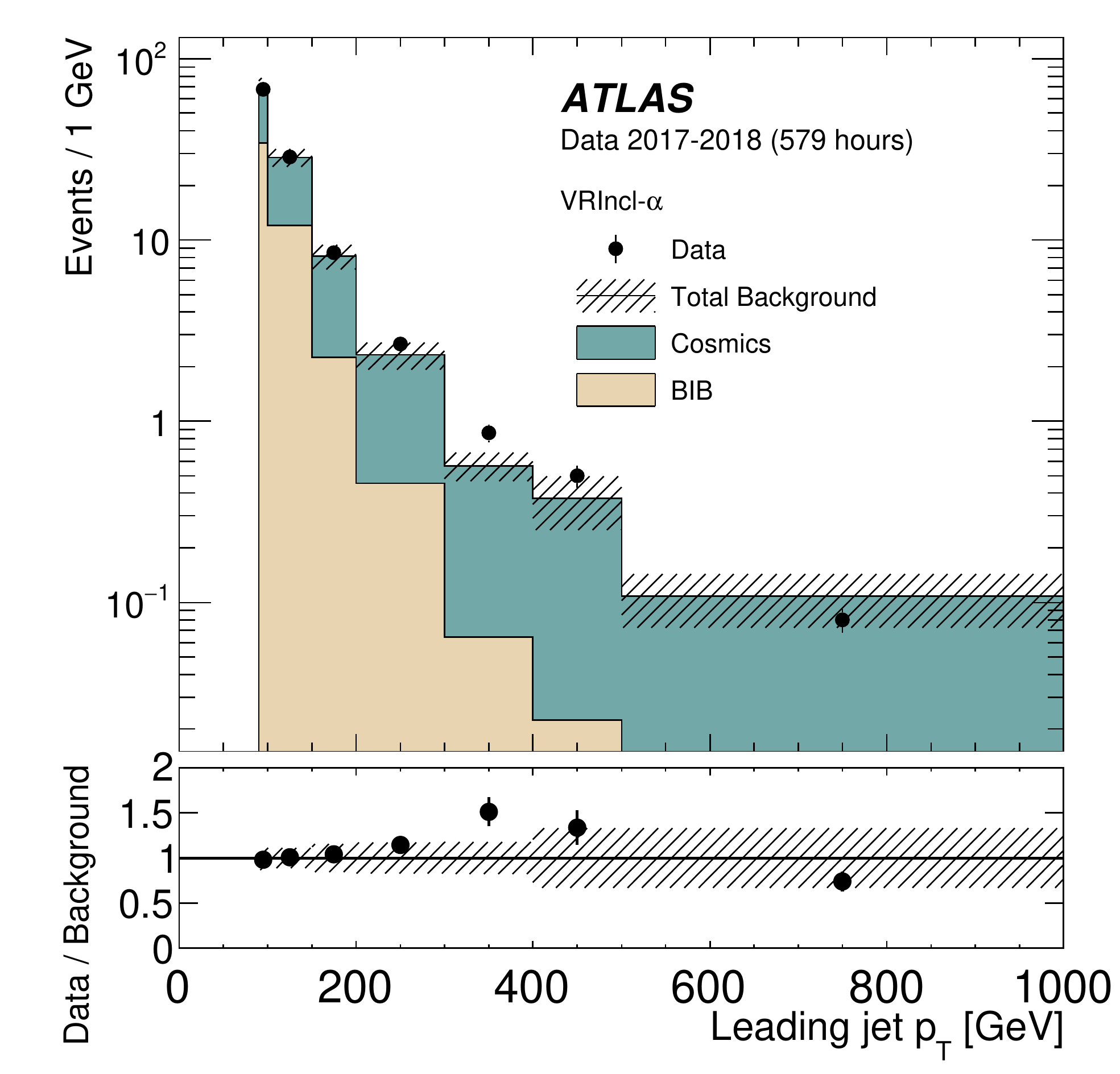}
\includegraphics[width=0.44\textwidth]{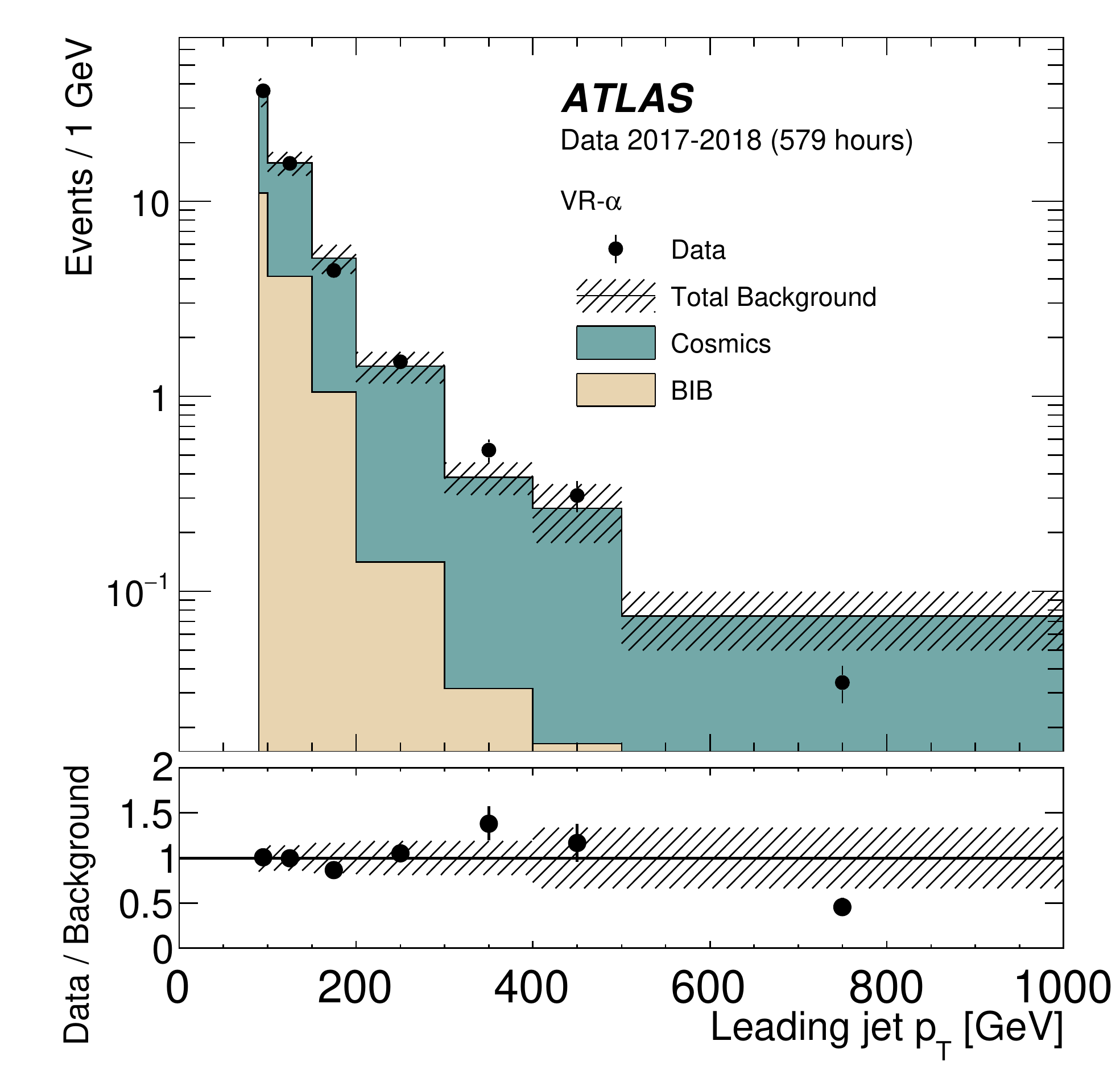}
\caption{The predicted leading-jet \pt\ distributions in VR-\widthphi\ (top),
VR-bib (middle) and VR-$\alpha$ (bottom), with the data from the 2017--2018 search sample overlaid.
Each distribution includes the low-\pt\ bins corresponding to the BIB NR associated with the VR.
Distributions are shown both for an $\eta$-inclusive selection (left) and for a central (right) selection.
The hatched band includes the statistical and systematic uncertainty of the background prediction.
The last bin includes the overflow.
The live time for the combined 2017 and 2018 empty-BX dataset is 579 hours.}
\label{fig:VRs}
\end{figure}
 
\FloatBarrier
 
\section{Systematic uncertainties}
\label{sec:syst}
 
The systematic uncertainties applied to the background estimates and the simulated signal samples are summarised in this section.
 
For the cosmics background, an uncertainty is assigned to the cosmics transfer factor to account for the limited number of events in the relevant regions in the cosmic sample.
An uncertainty is applied to the reweighting of the \pt\ distribution for the cosmics template, corresponding to 100\% of the impact of the reweighting.
An additional 20\% uncertainty is applied to account for differences between the shapes of the leading-jet \pt\ distribution in the cosmic sample and the cosmics CRs.
 
For the BIB prediction, an uncertainty is assigned to the BIB normalisation factor to account for the limited number of events in the relevant NR.
The uncertainties in the cosmics background template used to subtract the cosmics contribution from the BIB template are propagated to the BIB estimation.
In order to take into account differences between the leading-jet \pt\ distributions of BIB-jets in unpaired- and empty-BX data, a shape uncertainty is derived
using the observed shape difference in an inclusive region with $\alpha>0.2$.
This shape uncertainty is treated as uncorrelated between 2017 and 2018 data, owing to the different bunch schemes and sample sizes in the comparison.
 
The magnitude of each uncertainty contribution relative to the total predicted background in each jet \pt\ bin of SRC for 2017 and 2018 data is illustrated in Figure~\ref{fig:syst}.
The $\pt>500$~\GeV\ jet bin in each region dominates the analysis sensitivity, and the largest uncertainty contribution in this region is due to the shape uncertainties of the cosmics background prediction.
This includes the uncertainty due to both the reweighting procedure and the shape comparison with the cosmic sample.
Of these contributions, the reweighting uncertainty is the dominant contributor in this jet \pt\ range.
Uncertainties associated with the BIB prediction become more important in the lower jet-\pt\ bins,
where the relative contribution from BIB to the total background estimate is more significant.
For 2017 data, differences in the shape of the BIB-jet \pt\ distribution when comparing empty- and unpaired-BX data are the largest (second largest) source of uncertainty for the $300<\pt<500$~\GeV\ ($150<\pt<300$~\GeV) jet bin.
This is partly due to beam instabilities within the LHC, which necessitated frequent changes to the LHC filling scheme
(and therefore the bunch configuration) during 2017 data-taking operations.
The magnitude of the differences observed between BIB-jet \pt\ distributions in the search and BIB samples in 2017 are consequently larger than those observed in 2018 data.
 
For simulated signal samples, uncertainties are applied in order to take into account the non-projective and out-of-time nature of jets resulting from \rhadron\ decays.
To account for the impact on jet energy reconstruction of jets arriving out-of-time,
an uncertainty is derived from variations in the jet \pt\ spectrum of cosmic-ray-induced jets as a function of jet time-of-arrival.
Since the \pt\ spectrum for these jets should be independent of time-of-arrival,
the variation in \pt\ as a function of time is taken as a \pt-dependent uncertainty to account for biases as a result of late or early arrival of the jets relative to the LHC clock.
This procedure results in a \pt-dependent uncertainty, with an impact ranging from $5$\% to $50$\% in the highest jet-\pt\ bin.
To account for the impact of jet non-projectivity on the jet energy reconstruction, the variation in the leading-jet \pt\ distribution for cosmic-ray events as a function of $\phi$ is used.
Since cosmic-ray muons approach the detector in a downwards direction, by looking at the full detector $\phi$ range all projective possibilities are considered.
This results in an uncertainty of up to $12$\% in the highest jet-\pt\ bin.
For both the timing and the projection uncertainty, the maximum deviation upwards and downwards from the mean in each \pt\ bin is taken as an upwards and downwards uncertainty in the jet energy, respectively.
No further experimental systematic uncertainties are applied to the predicted signal yields.
Signal cross-sections are calculated to approximate next-to-next-to-leading order in the strong coupling constant,
adding the resummation of soft gluon emission at next-to-next-to-leading-logarithm accuracy (approximate NNLO+NNLL)~\cite{Beenakker:2016lwe,Beenakker:2014sma,Beenakker:2013mva,Beenakker:2011sf,Beenakker:2009ha,Kulesza:2009kq,Kulesza:2008jb,Beenakker:1996ch}.
The nominal cross-section and its uncertainty are derived using the PDF4LHC15\_mc PDF set, following the recommendations of Ref.~\cite{Butterworth:2015oua}.

\begin{figure}
\centering
\includegraphics[width=0.48\textwidth]{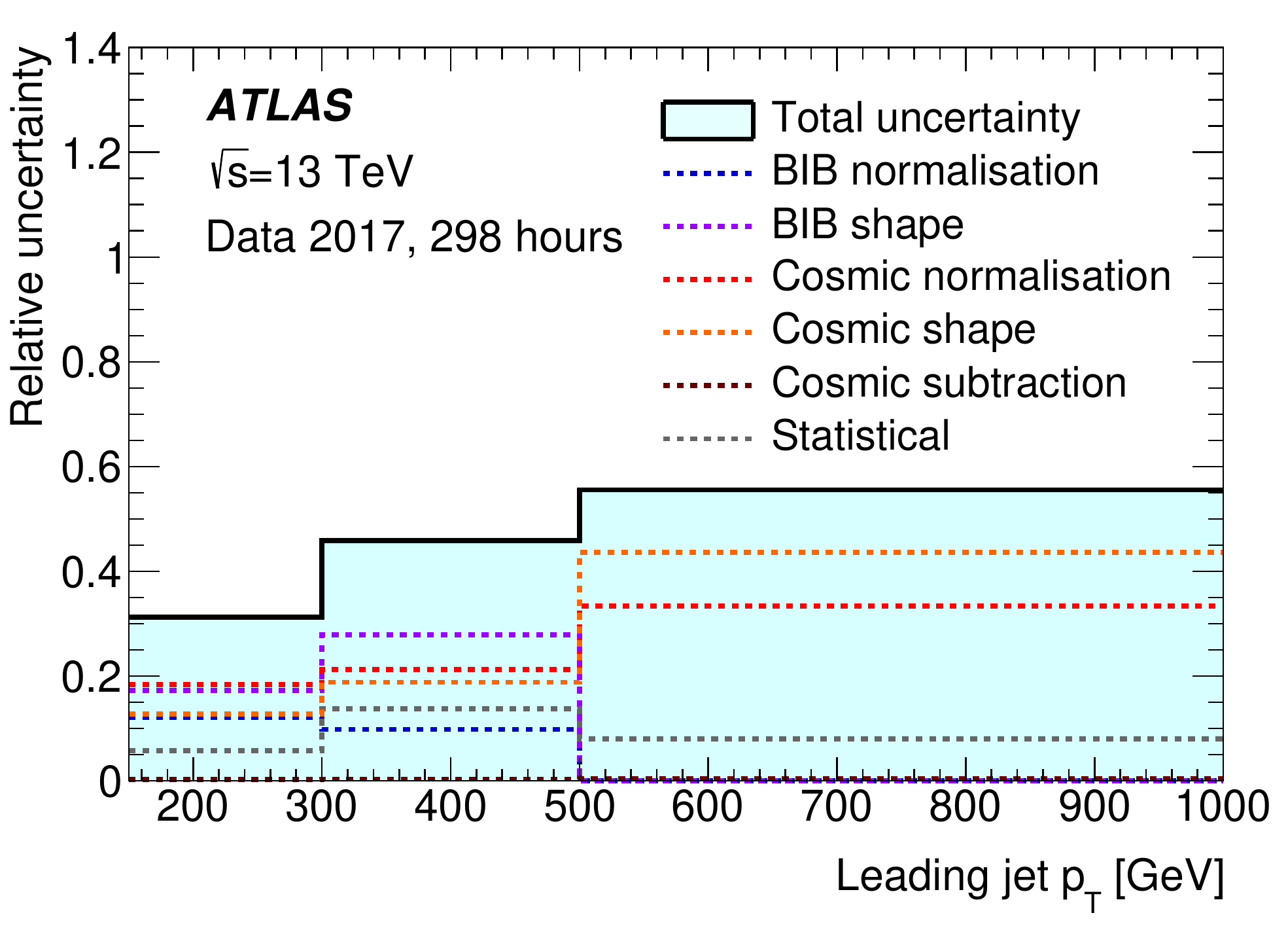}
\includegraphics[width=0.48\textwidth]{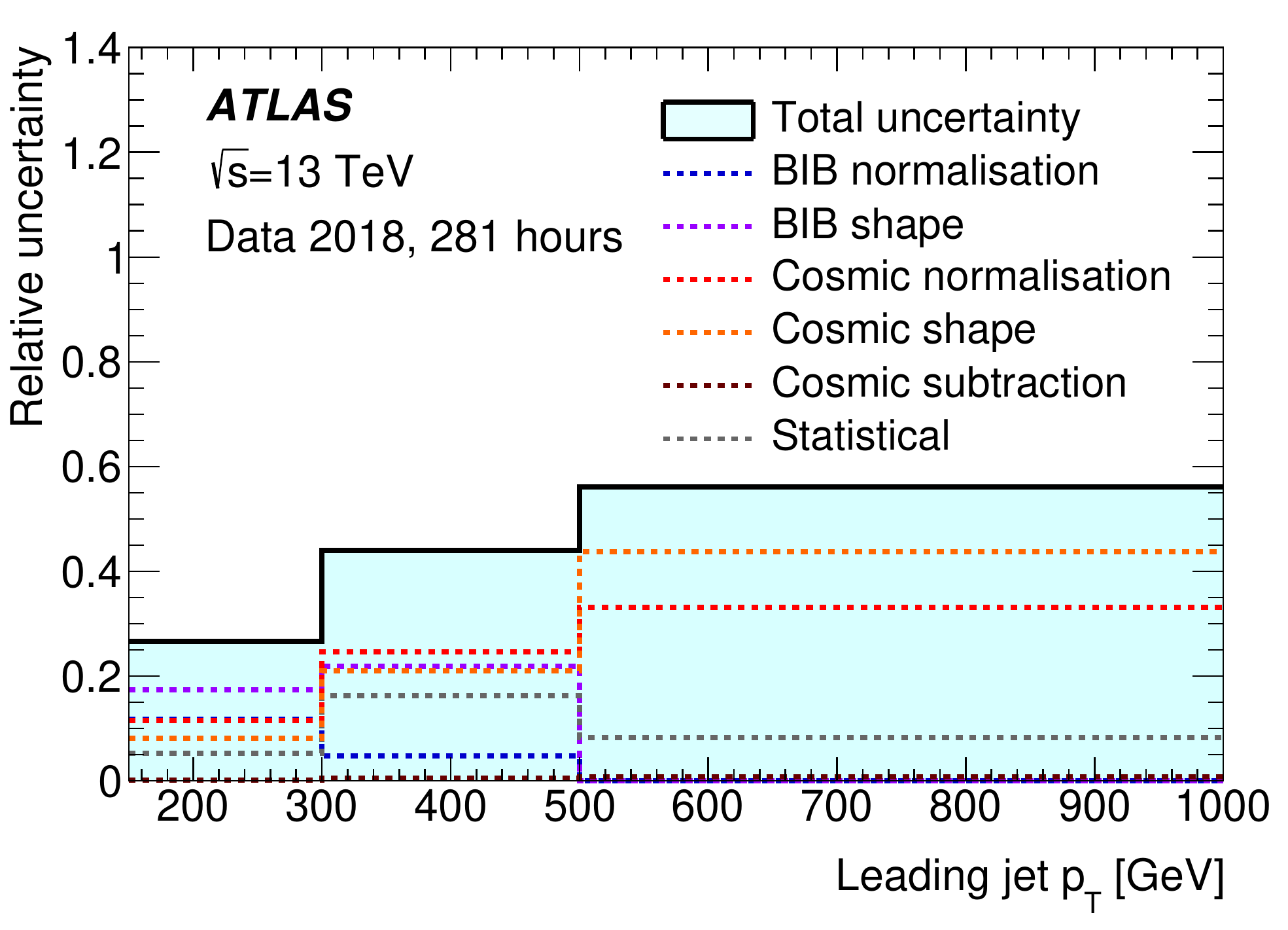}
\caption{Summary of the relative uncertainties in the total expected background prediction in each jet \pt\ bin of SRC for 2017 (left) and 2018 (right) data.
}
\label{fig:syst}
\end{figure}
 
\FloatBarrier
 
\section{Results}
\label{sec:result}

The integrated yields in the central and inclusive signal regions are compared with the expected background in Tables~\ref{tab:SRC_yields} and~\ref{tab:SRIncl_yields}, respectively, for 2017 and 2018 data.
The binned jet-\pt\ distributions are shown for these regions in Figure~\ref{fig:SRs}.
The central control and normalisation regions are used to provide the background estimates in SRC, and the inclusive control and normalisation regions are used to provide the predicted yields in SRIncl.
 
\begin{table}
\centering
\caption{Breakdown of the expected and observed data yields in SRC in 2017 and 2018, integrated over the jet \pt\ spectrum.
The quoted uncertainties include statistical and systematic contributions.
The expected event yields from two example signal models with a gluino mass of $1.4$~\TeV\ and neutralino masses of $100$~\GeV\ and $900$~\GeV\ are also included,
where the yields are calculated assuming a gluino lifetime  between about $10^{-5}$ and $10^3$~s (within the range of the live-fraction plateau in Figure~\ref{fig:LiveFractionRun2}). }
\vspace{0.2cm}
\small
\begin{tabular*}{\textwidth}{@{\extracolsep{\fill}}lll}
\toprule
\textbf{Central signal regions}     & SRC (2017 data)   & SRC (2018 data) \\
\midrule
\midrule
Observed events                          & ~~92                & 100\\
\midrule
Total expected background events         & ~~$88 \pm 28$ & $119 \pm 32$  \\
\midrule
Beam-induced background events           & ~~$37 \pm 23$ & ~~$72 \pm 29$   \\
Cosmic-ray-induced background events         & ~~$51 \pm 21$ & ~~$47 \pm 19$    \\
\midrule
 
$m(\tilde{g},\tilde{\chi}_{1}^{0})=(1400, 100)$~\GeV  & ~~~~5  &  ~~~~6   \\ 
$m(\tilde{g},\tilde{\chi}_{1}^{0})=(1400, 900)$~\GeV  & ~~~~5  &  ~~~~6   \\ 
\bottomrule
\end{tabular*}
\label{tab:SRC_yields}
\end{table}

\begin{table}
\centering
\caption{Breakdown of the expected and observed data yields in SRIncl in 2017 and 2018, integrated over the jet \pt\ spectrum.
The quoted uncertainties include statistical and systematic contributions.
The expected event yields from two example signal models with a gluino mass of $1.4$~\TeV\ and neutralino masses of $100$~\GeV\ and $900$~\GeV\ are also included,
where the yields are calculated assuming a gluino lifetime between about $10^{-5}$ and $10^3$~s (within the range of the live-fraction plateau in Figure~\ref{fig:LiveFractionRun2}). }
\vspace{0.2cm}
\small
\begin{tabular*}{\textwidth}{@{\extracolsep{\fill}}lll}
\toprule
\textbf{Inclusive signal regions}   &        SRIncl (2017 data)   & SRIncl (2018 data) \\
\midrule
Observed events                           & $239$               & $221$\\
\midrule
Total expected background events         & $167 \pm 48$   &  $208 \pm 50$\\
\midrule
Beam-induced background events           & ~~$93 \pm 42$    & $139 \pm 45$\\
Cosmic-ray-induced background events        & ~~$74 \pm 30$    & ~~$69 \pm 28$ \\
\midrule
 
$m(\tilde{g},\tilde{\chi}_{1}^{0})=(1400, 100)$~\GeV  & ~~~~7  &  ~~~~9   \\ 
$m(\tilde{g},\tilde{\chi}_{1}^{0})=(1400, 900)$~\GeV  & ~~~~7  &  ~~~~8   \\ 
\bottomrule
\end{tabular*}
\label{tab:SRIncl_yields}
\end{table}

\begin{figure}
\centering
\includegraphics[width=0.45\textwidth]{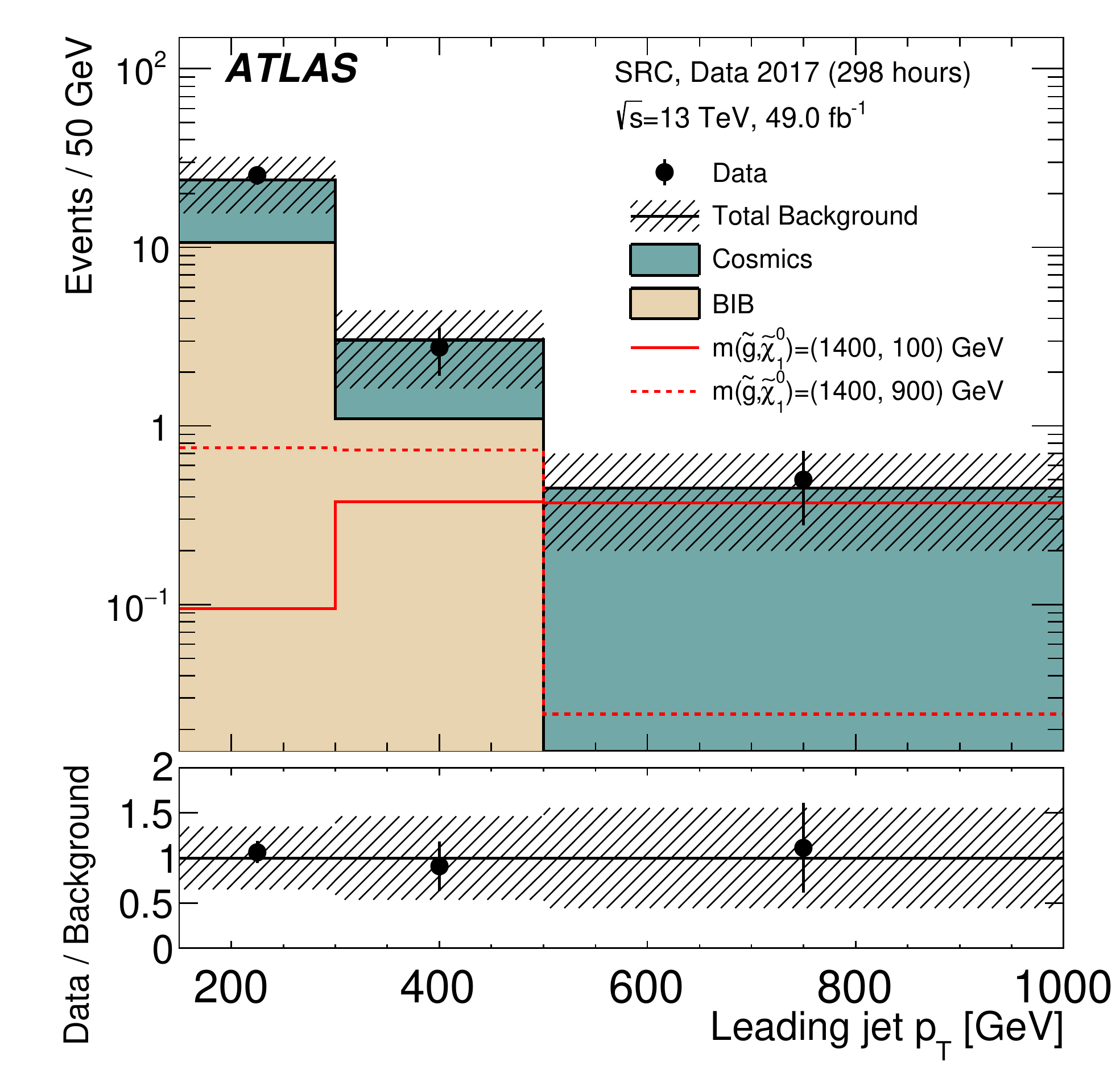}
\includegraphics[width=0.45\textwidth]{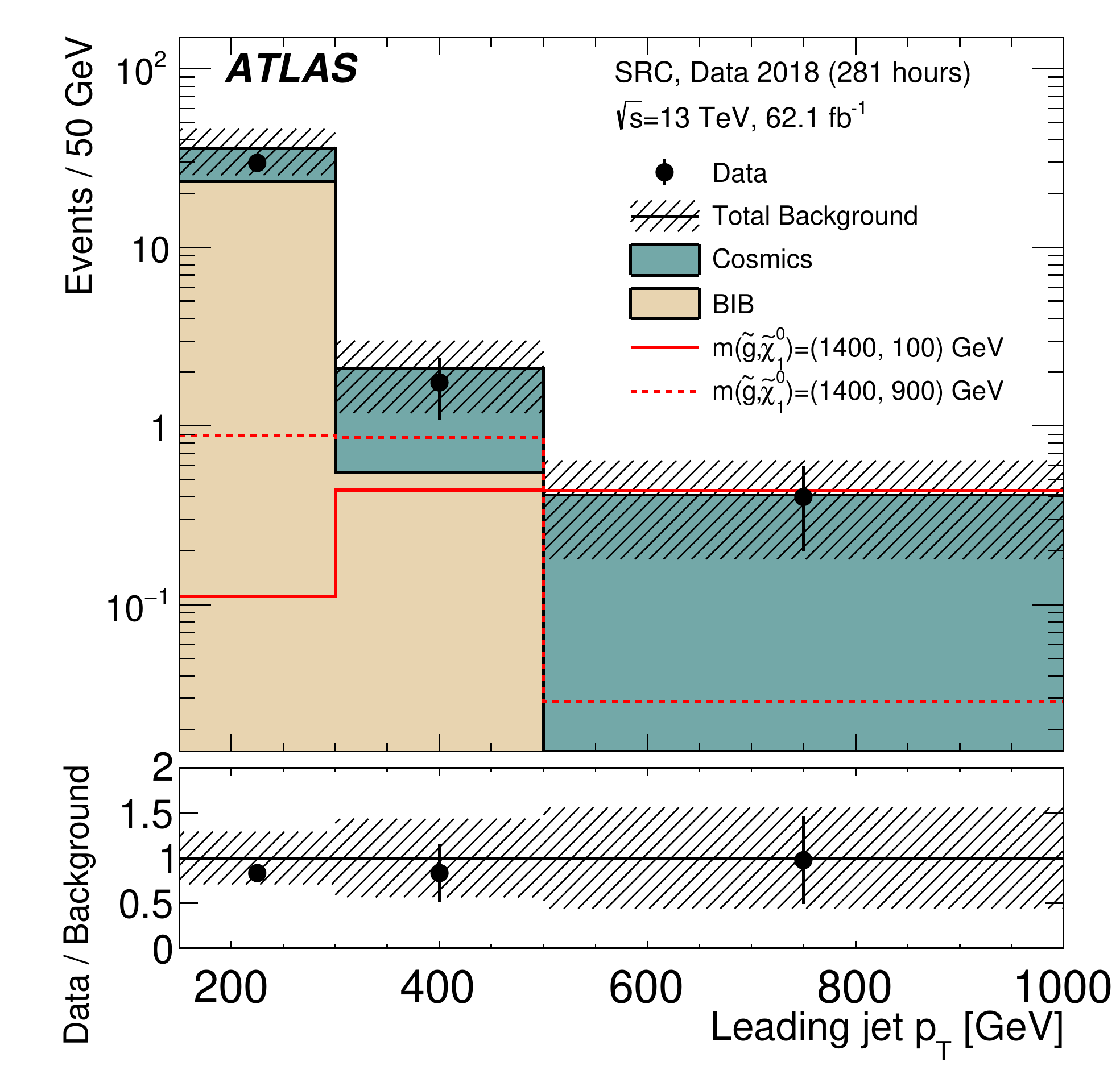}
\includegraphics[width=0.45\textwidth]{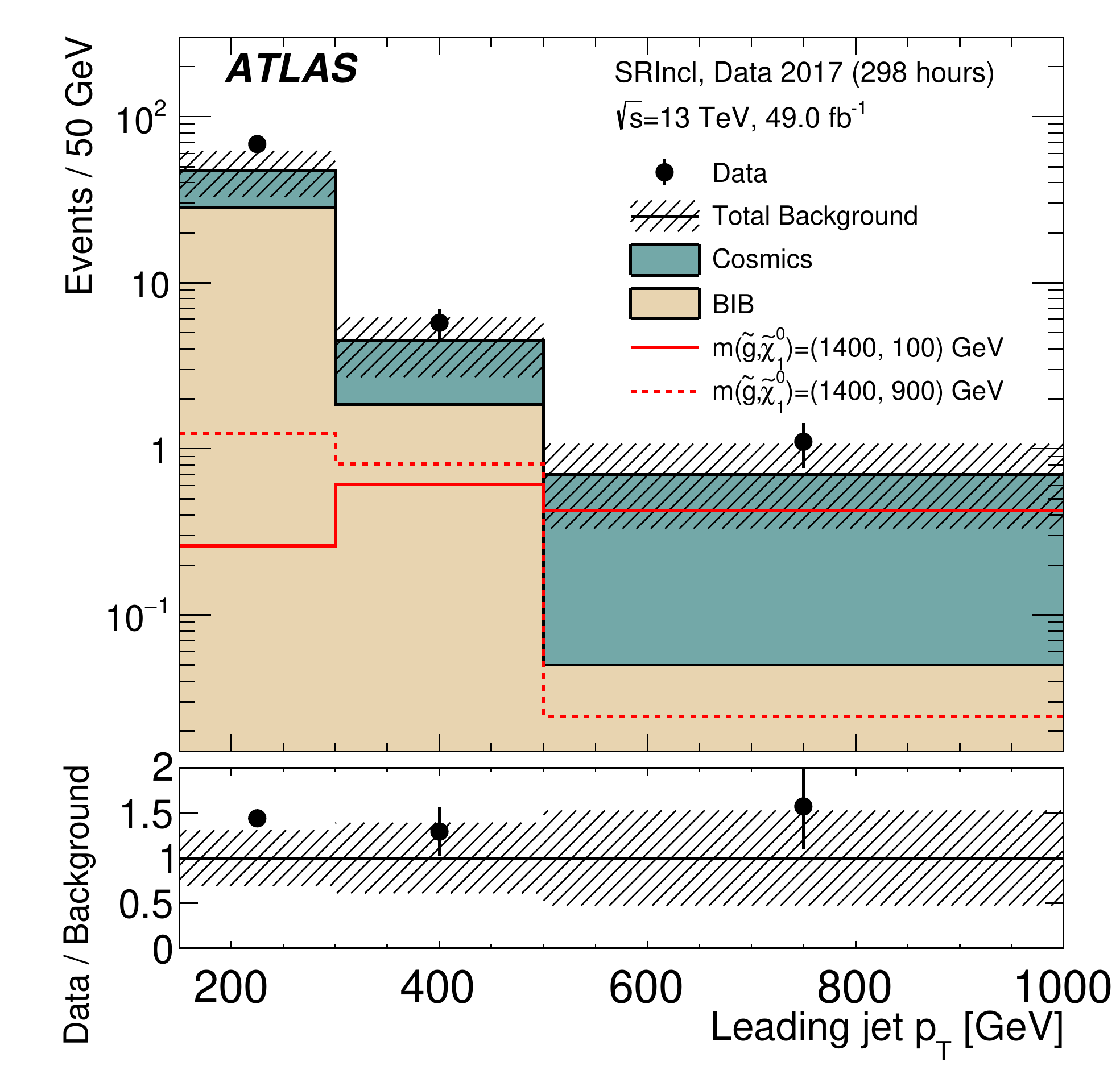}
\includegraphics[width=0.45\textwidth]{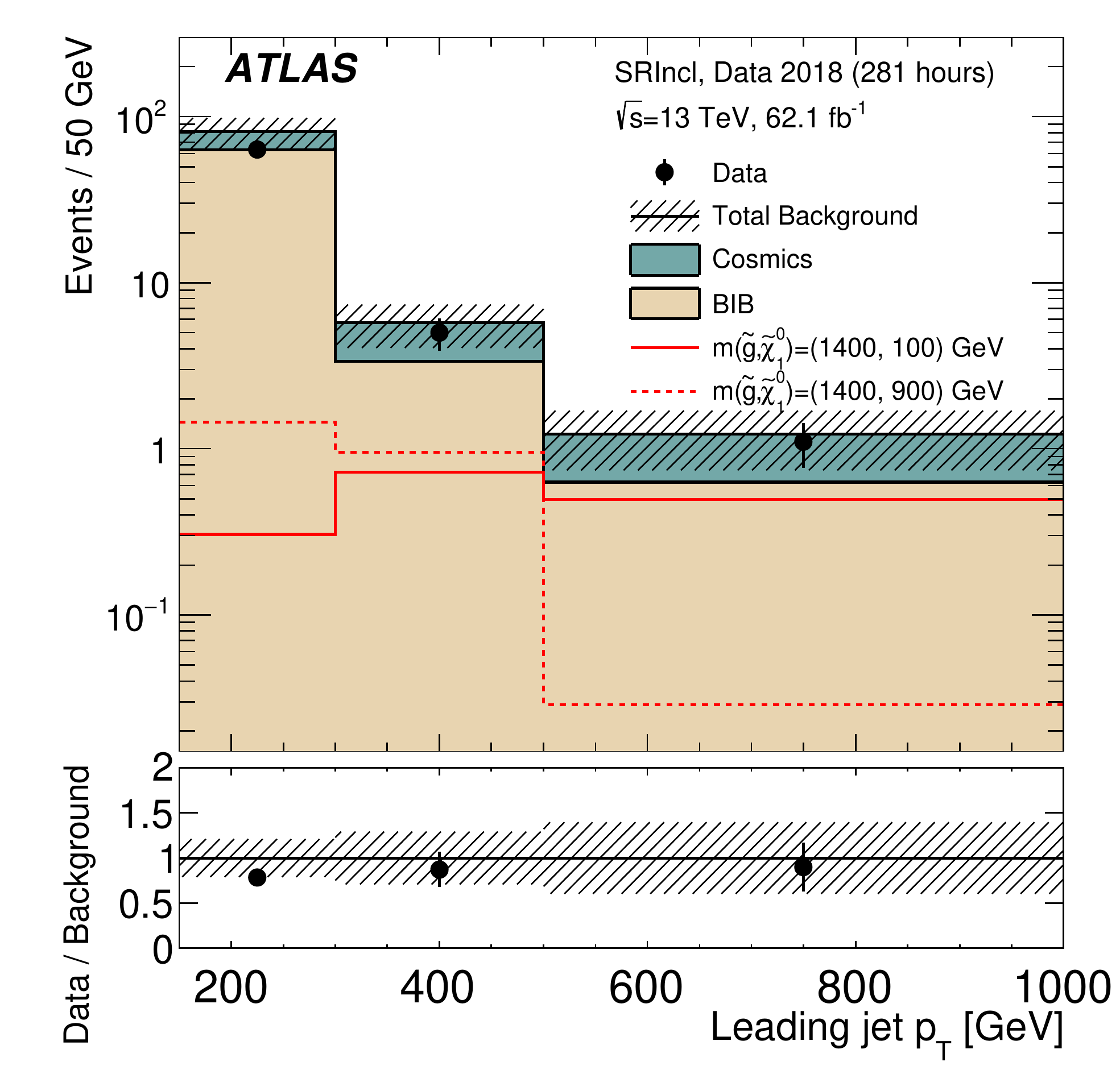}
\caption{Observed and expected leading-jet \pt\ distributions, with the bin boundaries considered for the interpretation, in 2017 data (left) and 2018 data (right) in SRC (top) and SRIncl (bottom).
All statistical and systematic uncertainties of the expected background are included in the hatched band.
The last bin contains the overflow.
Two example signal models with a gluino mass of $1.4$~\TeV\ and neutralino masses of $100$~\GeV\ and $900$~\GeV\ are overlaid,
where the yields are calculated assuming a gluino lifetime within the range of the live-fraction plateau. }
\label{fig:SRs}
\end{figure}
 
Assumptions about the mass difference $\deltam$ impact the point at which the jet \pt\ distributions of various signal models are expected to peak.
Any potential excess of events must therefore be searched for across the range of the leading-jet \pt\ distribution.
To do this the three exclusive jet \pt\ bins in each SR are made inclusive to form the discovery regions, used to make model-independent statements about the possible presence of new physics.
Because the signal cross-section sensitivity of the 2018 dataset is expected to be roughly 20\% larger than the 2017 dataset,
only 2018 data is used to make these model-independent statements.
The results in the central and inclusive DRs are summarised in Tables~\ref{tab:DRC} and~\ref{tab:DRIncl}, respectively.
The data are consistent with the expected background across the full jet \pt\ range.
 
Model-independent upper limits at $95$\% confidence level (CL) on the number of events ($S^{95}$) that could be attributed to BSM physics processes are derived using the CL$_{\mathrm{S}}$ prescription~\cite{Read:2002hq},
implemented using HistFitter~\cite{Baak:2014wma}.
This procedure is carried out using the DRs summarised in Table~\ref{tab:SRs}, and neglects potential signal contamination in the CRs.
Pseudo-experiments are used to set these upper limits.
The expected ($S_\text{exp}^{95}$) and observed ($S_\text{obs}^{95}$) upper limits on the number of possible BSM events are provided in Tables~\ref{tab:DRC} and~\ref{tab:DRIncl}.
The $p$-values, which represent the probability of cosmic-ray- or beam-induced backgrounds alone to fluctuate to the observed number of events or higher, are also provided.

\begin{table}
\centering
\caption{Central (jet $|\eta|<0.8$) discovery regions in 2018 data. The lower lines of the table show the 95\% CL upper limits on the number of signal events ($S_\text{obs}^{95}$ ), the 95\% CL upper limit on the number of signal events given the expected number (and $\pm 1\sigma$ excursions of the expectation) of background events ($S_\text{exp}^{95}$), the $CL_B$ value, i.e.~the confidence level observed for the background-only hypothesis, and the discovery $p$-value ($p(s = 0)$). The $p$-value is reported as 0.5 if the observed yield is smaller than the predicted. }
\vspace{0.2cm}
\small
\begin{tabular*}{\textwidth}{@{\extracolsep{\fill}}llll}
\toprule
\textbf{Central discovery regions}           & Jet $\pt>150$~\GeV     & Jet $\pt>300$~\GeV     & Jet $\pt>500$~\GeV              \\
\midrule
Observed events                          & 100                    & ~~11              & ~~~~4                    \\
\midrule
Total expected background events         & $119 \pm 32$     & ~~$12.5 \pm 5.3$          & ~~~~$4.1 \pm 2.3$              \\
\midrule
Beam-induced background events           & ~~$72 \pm 29$      & ~~~~$2.2 \pm 2.1$          & ~~~~$0.00_{-0.00}^{+0.05}$              \\
Cosmic-ray-induced background events         & ~~$47 \pm 19$      & ~~$10.3 \pm 5.0$          & ~~~~$4.1 \pm 2.3$              \\
\midrule
$S_\text{exp}^{95}$  & ~~$ { 52.8 }^{ +14.6 }_{ -13.7 }$ & ~~$ { 14.3 }^{ +1.4 }_{ -2.9 }$ & ~~~~$ { 7.8 }^{ +4.6 }_{ -0.2 }$ \\
$S_\text{obs}^{95}$  & ~~$45.3$ & ~~$14.1$ & ~~~~$7.9$ \\
$CL_{B}$ & ~~~~$0.30$ & ~~~~$0.45$ & ~~~~$0.56$ \\
$p(s=0)$ ($Z$) & ~~~~$0.5$ $(0)$ & ~~~~$0.5$ $(0)$ & ~~~~$0.5$ $(0)$ \\
\bottomrule
\end{tabular*}
\label{tab:DRC}
\end{table}
 
\begin{table}
\centering
\caption{Inclusive (jet $|\eta|<2.4$) discovery regions in 2018 data. The lower lines of the table show the 95\% CL upper limits on the number of signal events ($S_\text{obs}^{95}$ ), the 95\% CL upper limit on the number of signal events given the expected number (and $\pm 1\sigma$ excursions of the expectation) of background events ($S_\text{exp}^{95}$), the $CL_B$ value, i.e.~the confidence level observed for the background-only hypothesis, and the discovery $p$-value ($p(s = 0)$).}
\vspace{0.2cm}
\small
\begin{tabular*}{\textwidth}{@{\extracolsep{\fill}}llll}
\toprule
\textbf{Inclusive discovery regions}   & Jet $\pt>150$~\GeV     & Jet $\pt>300$~\GeV     & Jet $\pt>500$~\GeV       \\
\midrule
Observed events                             & 221              & ~~31              & ~~11                    \\
\midrule
Total expected background events            & $208 \pm 50$          & ~~$19.7 \pm 7.6$          & ~~~~$6.2 \pm 3.4$              \\
\midrule
Beam-induced background events              & $139 \pm 45$          & ~~~~$4.5 \pm 2.3$          & ~~~~$0.38 \pm 0.23$              \\
Cosmic-ray-induced background events        & ~~$69 \pm 28$          & ~~$15.2 \pm 7.2$          & ~~~~$5.8 \pm 3.4$              \\
\midrule
$S_\text{exp}^{95}$  & $ ~~{ 97 }^{ +35 }_{ -26 }$ & $ ~~{ 19.6 }^{ +8.3 }_{ -5.7 }$ & ~~$ { 12.7 }^{ +2.3 }_{ -4.8 }$ \\
$S_\text{obs}^{95}$  & $104$ & ~~$29.3$  & ~~$15.5$ \\
$CL_{B}$ & ~~~~$0.62$ & ~~~~$0.89$ & ~~~~$0.87$ \\
$p(s=0)$ ($Z$) & ~~~~$ 0.27$~$(0.63)$ & $ ~~~~0.12$~$(1.17)$ & ~~~~$ 0.14$~$(1.06)$ \\
\bottomrule
\end{tabular*}
\label{tab:DRIncl}
\end{table}

Exclusion limits are set at 95\% CL as a function of the gluino mass and mean proper lifetime, for different gluino--neutralino mass-splitting hypotheses, and can be found in Figure~\ref{fig:limits}.
For the signal models considered in this interpretation the central signal regions dominate the sensitivity, so the limits presented here are based on the results in SRC alone.
SRC in this context is implemented in the fit as two separate three-bin regions, with one region for 2017 data and the other for 2018 data (these regions are shown in the top panel of Figure~\ref{fig:SRs}).
The normalisation region, NRC-bib, is included as two separate single-bin regions, again with one region per year.
The hypothesis tests include the expected signal yield and its associated uncertainties in the normalisation regions and SRs,
so potential signal contamination in NRC-bib is taken into account in the exclusion fit.
 
Gluinos with $\taugl=100$~ns are excluded up to a mass of $1.0$~\TeV\ for a neutralino mass of $100$~\GeV,
with this lifetime corresponding to the minimum time between the last $pp$ collision and the first empty BX used in this analysis.
The live-fraction efficiency plateau is reached for $\taugl$ approximately equal to a quarter of the time taken for a single full LHC revolution.
While on the plateau, gluino masses of up to $1.4$~\TeV\ are excluded for lifetimes up to a few hours ($10^{-5}$ to $10^3$~s), after which point the signal efficiency decreases.
Gluino masses of $1.0$~\TeV\ are excluded for lifetimes of up to $10^{7}$~s.
The sensitivity is significantly weaker when smaller mass splittings are considered ($\deltam = 100$~\GeV) due to a reduced selection acceptance. In these cases, the sensitivity is driven by the expected signal yields in NRC-bib. Limits on the signal production cross-section are also shown as a function of gluino mass in Figure~\ref{fig:xslimits}, assuming a gluino lifetime of $\taugl=1$~ms.
 
\begin{figure}
\centering
\includegraphics[width=0.8\textwidth]{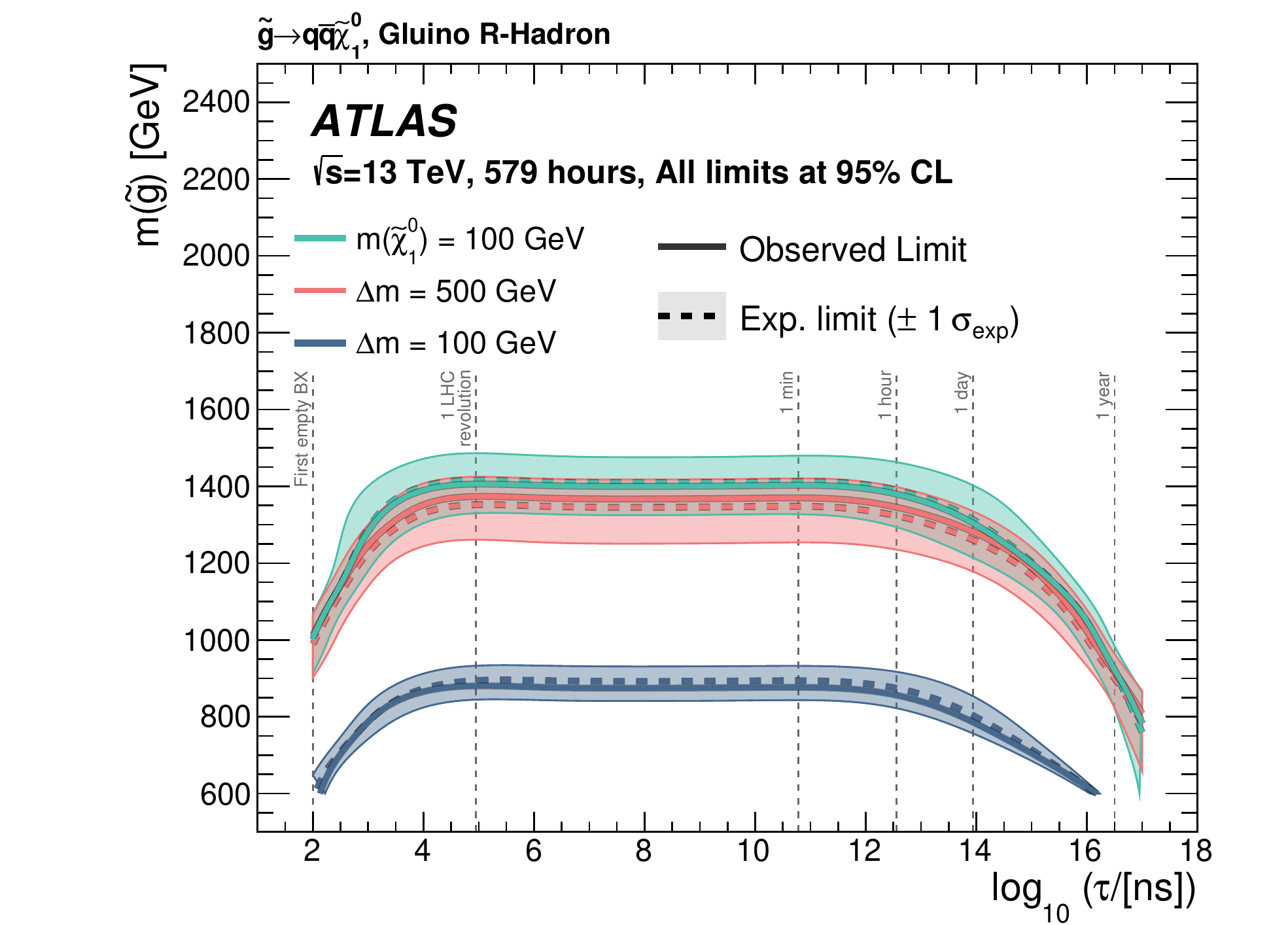}
\caption{Expected (dashed lines) and observed (solid lines) exclusion limits at 95\% CL using the 2017 and 2018 datasets in the central signal regions. The limits are shown as a function of gluino mass and $\taugl$. The shaded coloured bands represent the $\pm 1 \sigma_{\mathrm{exp}}$ variations from systematic and statistical uncertainties in the expected yields. The different sets of colours represent the limits for different assumptions about the gluino--neutralino mass splitting.}
\label{fig:limits}
\end{figure}
 
\begin{figure}
\centering
\includegraphics[width=0.8\textwidth]{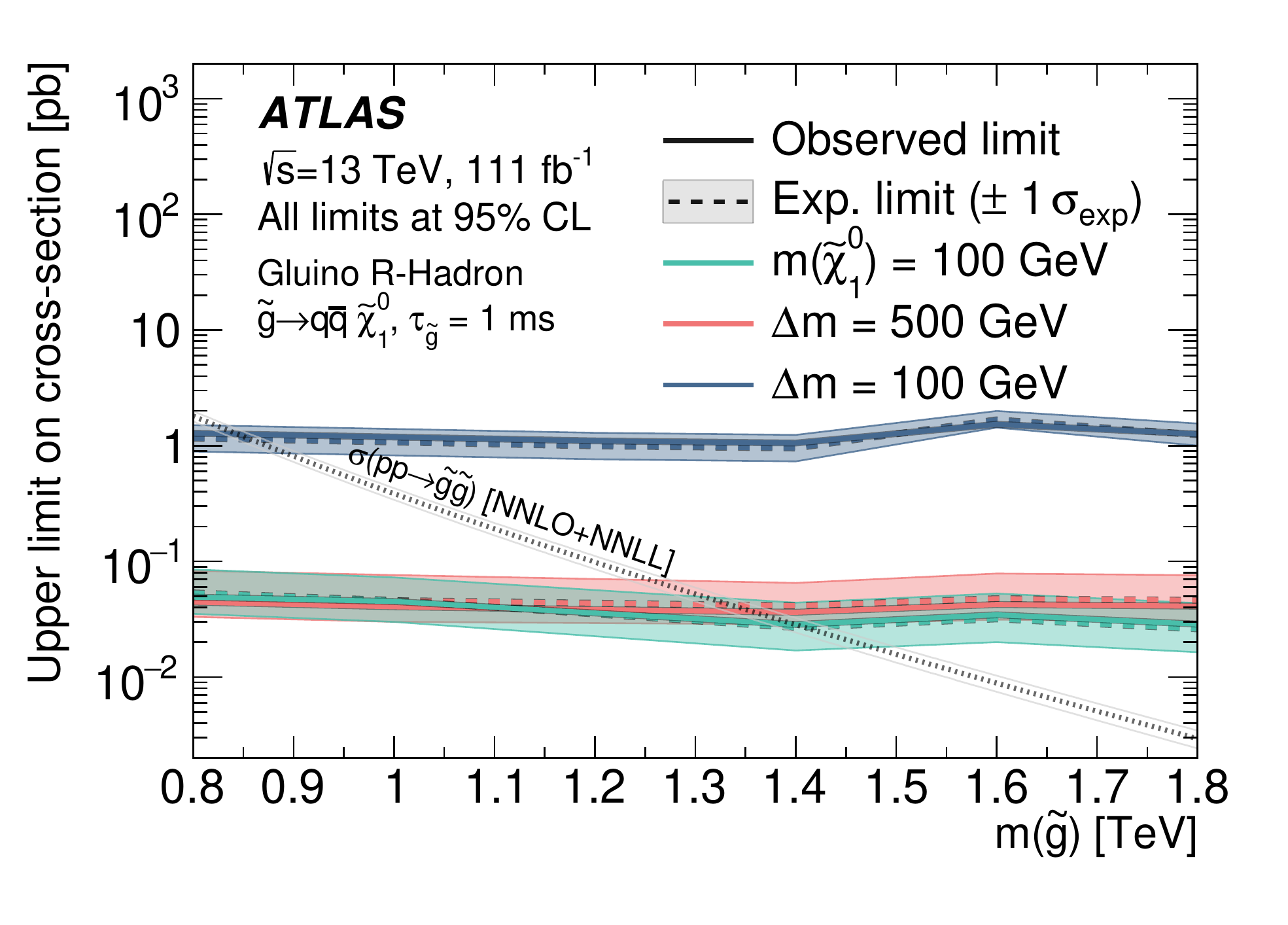}
\caption{Expected (dashed lines) and observed (solid lines) signal cross-section exclusion limits at 95\% CL using the 2017 and 2018 datasets in the central signal regions. The cross-section limits are shown as a function of gluino mass. The shaded coloured bands represent the $\pm 1 \sigma_{\mathrm{exp}}$ variations from systematic and statistical uncertainties in the expected yields. The different sets of colours represent the limits for different assumptions about the gluino--neutralino mass splitting. The grey dotted line and band represent the NNLO+NNLL theory prediction and uncertainties in the production cross-section, respectively.}
\label{fig:xslimits}
\end{figure}

\FloatBarrier
 
\section{Conclusion}
\label{sec:conclusion}
 
This paper presents a search for long-lived particles that have come to rest within the ATLAS detector at the LHC,
using data collected in empty bunch crossings during 2017 and 2018 $\sqrt{s}=13$~\TeV\ $pp$ collision data-taking. The data are found to be consistent with the background prediction,
and the results are interpreted in the context of simplified models of gluino pair-production,
where the gluinos form \rhadrons.
Gluino masses up to 1.4~\TeV\ are excluded for gluino lifetimes of $10^{-5}$ to $10^3$~s, assuming a neutralino mass of $100$~\GeV.
This analysis represents the first search for stopped long-lived particles at ATLAS using $\sqrt{s}=13$~\TeV\ $pp$ collision data,
significantly expands the limits on such signatures given by previous ATLAS analyses, and includes new methodologies for the estimation of non-collision background processes.

\section*{Acknowledgements}
 

We thank CERN for the very successful operation of the LHC, as well as the
support staff from our institutions without whom ATLAS could not be
operated efficiently.
 
We acknowledge the support of ANPCyT, Argentina; YerPhI, Armenia; ARC, Australia; BMWFW and FWF, Austria; ANAS, Azerbaijan; SSTC, Belarus; CNPq and FAPESP, Brazil; NSERC, NRC and CFI, Canada; CERN; ANID, Chile; CAS, MOST and NSFC, China; COLCIENCIAS, Colombia; MSMT CR, MPO CR and VSC CR, Czech Republic; DNRF and DNSRC, Denmark; IN2P3-CNRS and CEA-DRF/IRFU, France; SRNSFG, Georgia; BMBF, HGF and MPG, Germany; GSRT, Greece; RGC and Hong Kong SAR, China; ISF and Benoziyo Center, Israel; INFN, Italy; MEXT and JSPS, Japan; CNRST, Morocco; NWO, Netherlands; RCN, Norway; MNiSW and NCN, Poland; FCT, Portugal; MNE/IFA, Romania; JINR; MES of Russia and NRC KI, Russian Federation; MESTD, Serbia; MSSR, Slovakia; ARRS and MIZ\v{S}, Slovenia; DST/NRF, South Africa; MICINN, Spain; SRC and Wallenberg Foundation, Sweden; SERI, SNSF and Cantons of Bern and Geneva, Switzerland; MOST, Taiwan; TAEK, Turkey; STFC, United Kingdom; DOE and NSF, United States of America. In addition, individual groups and members have received support from BCKDF, CANARIE, Compute Canada, CRC and IVADO, Canada; Beijing Municipal Science \& Technology Commission, China; COST, ERC, ERDF, Horizon 2020 and Marie Sk{\l}odowska-Curie Actions, European Union; Investissements d'Avenir Labex, Investissements d'Avenir Idex and ANR, France; DFG and AvH Foundation, Germany; Herakleitos, Thales and Aristeia programmes co-financed by EU-ESF and the Greek NSRF, Greece; BSF-NSF and GIF, Israel; La Caixa Banking Foundation, CERCA Programme Generalitat de Catalunya and PROMETEO and GenT Programmes Generalitat Valenciana, Spain; G\"{o}ran Gustafssons Stiftelse, Sweden; The Royal Society and Leverhulme Trust, United Kingdom.
 
The crucial computing support from all WLCG partners is acknowledged gratefully, in particular from CERN, the ATLAS Tier-1 facilities at TRIUMF (Canada), NDGF (Denmark, Norway, Sweden), CC-IN2P3 (France), KIT/GridKA (Germany), INFN-CNAF (Italy), NL-T1 (Netherlands), PIC (Spain), ASGC (Taiwan), RAL (UK) and BNL (USA), the Tier-2 facilities worldwide and large non-WLCG resource providers. Major contributors of computing resources are listed in Ref.~\cite{ATL-SOFT-PUB-2020-001}.
 
 
\clearpage
 
\printbibliography
 
\FloatBarrier

\clearpage 
 
\begin{flushleft}
\hypersetup{urlcolor=black}
{\Large The ATLAS Collaboration}

\bigskip

\AtlasOrcid[0000-0002-6665-4934]{G.~Aad}$^\textrm{\scriptsize 101}$,    
\AtlasOrcid[0000-0002-5888-2734]{B.~Abbott}$^\textrm{\scriptsize 128}$,    
\AtlasOrcid[0000-0002-7248-3203]{D.C.~Abbott}$^\textrm{\scriptsize 102}$,    
\AtlasOrcid[0000-0002-2788-3822]{A.~Abed~Abud}$^\textrm{\scriptsize 36}$,    
\AtlasOrcid[0000-0002-1002-1652]{K.~Abeling}$^\textrm{\scriptsize 53}$,    
\AtlasOrcid[0000-0002-2987-4006]{D.K.~Abhayasinghe}$^\textrm{\scriptsize 93}$,    
\AtlasOrcid[0000-0002-8496-9294]{S.H.~Abidi}$^\textrm{\scriptsize 29}$,    
\AtlasOrcid[0000-0002-8279-9324]{O.S.~AbouZeid}$^\textrm{\scriptsize 40}$,    
\AtlasOrcid{N.L.~Abraham}$^\textrm{\scriptsize 156}$,    
\AtlasOrcid[0000-0001-5329-6640]{H.~Abramowicz}$^\textrm{\scriptsize 161}$
% ...
\bigskip
\\

% ...
$^{29}$Physics Department, Brookhaven National Laboratory, Upton NY; United States of America.\\
% ...
$^{36}$CERN, Geneva; Switzerland.\\
% ...
$^{40}$Niels Bohr Institute, University of Copenhagen, Copenhagen; Denmark.\\
% ...
$^{53}$II. Physikalisches Institut, Georg-August-Universit\"{a}t G\"ottingen, G\"ottingen; Germany.\\
% ...
$^{93}$Department of Physics, Royal Holloway University of London, Egham; United Kingdom.\\
% ...
$^{101}$CPPM, Aix-Marseille Universit\'e, CNRS/IN2P3, Marseille; France.\\
$^{102}$Department of Physics, University of Massachusetts, Amherst MA; United States of America.\\
% ...
$^{128}$Homer L. Dodge Department of Physics and Astronomy, University of Oklahoma, Norman OK; United States of America.\\
% ...
$^{156}$Department of Physics and Astronomy, University of Sussex, Brighton; United Kingdom.\\
% ...
$^{161}$Raymond and Beverly Sackler School of Physics and Astronomy, Tel Aviv University, Tel Aviv; Israel.\\
% ...

\end{flushleft}


\end{document}